\documentclass[structabstract]{aa}  
\usepackage{graphicx}
\usepackage{amsmath,amssymb}
\usepackage{longtable,lscape}
\usepackage{txfonts}
\begin{document}

\def\Arrow{\mathop{\longrightarrow}\limits}
\def\Harpoons{\mathop{\rightleftharpoons}\limits}

   \title{$^{13}$CO and C$^{18}$O $J=2-1$ mapping of the environment of the Class 0 protostellar core SMM 3 in Orion B9\thanks{This publication is based on data acquired with the Atacama Pathfinder EXperiment (APEX) under programme 088.F-9311A. APEX is a collaboration between the Max-Planck-Institut f\"{u}r Radioastronomie, the European Southern Observatory, and the Onsala Space Observatory.}}

   \author{O. Miettinen}

 \offprints{O. Miettinen}

   \institute{Department of Physics, P.O. Box 64, FI-00014 University of Helsinki, Finland\\ \email{oskari.miettinen@helsinki.fi}}

   \date{Received ; accepted}

\authorrunning{Miettinen}
\titlerunning{$^{13}$CO and C$^{18}$O mapping of Ori B9--SMM 3}

  \abstract
   {Observations of molecular spectral lines provide information on the gas
kinematics and chemistry of star-forming regions.}
   {We attempt to achieve a better understanding of the gas distribution and 
velocity field around the deeply embedded Class 0 protostar SMM 3 in the Orion 
B9 star-forming region.}
   {Using the APEX 12-m telescope, we mapped the line emission from the 
$J=2-1$ rotational transition of two CO isotopologues, $^{13}$CO and 
C$^{18}$O, over a $4\arcmin \times 4\arcmin$ region around Orion B9/SMM 3.} 
   {Both the $^{13}$CO and C$^{18}$O lines exhibit two well separated velocity 
components at about 1.3 and 8.7 km~s$^{-1}$. The emission of both CO 
isotopologues is more widely distributed than the submillimetre dust continuum 
emission as probed by LABOCA. The LABOCA 870-$\mu$m peak position of SMM 3 
is devoid of strong CO isotopologue emission, which is consistent with our 
earlier detection of strong CO depletion in the source. No signatures of a 
large-scale outflow were found towards SMM 3. The $^{13}$CO and C$^{18}$O 
emission seen at $\sim1.3$ km~s$^{-1}$ is concentrated into a single 
clump-like feature at the eastern part of the map. The peak H$_2$ column 
density towards a C$^{18}$O maximum of the low-velocity component is 
estimated to be $\sim10^{22}$ cm$^{-2}$. 
A velocity gradient was found across both the $^{13}$CO and C$^{18}$O maps. 
Interestingly, SMM 3 lies on the border of this velocity gradient.}
   {The $^{13}$CO and C$^{18}$O emission at $\sim1.3$ km~s$^{-1}$ is likely to 
originate from the ``low-velocity part'' of Orion B. Our analysis suggests 
that it contains high density gas ($\sim10^{22}$ H$_2$ molecules per cm$^2$), 
which conforms to our earlier detection of deuterated species at similarly low 
radial velocities. Higher-resolution observations would be needed to clarify 
the outflow activity of SMM 3. The sharp velocity gradient in the region might 
represent a shock front resulting from the feedback from the nearby 
expanding \ion{H}{ii} region NGC 2024. The formation of SMM 3, and possibly 
of the other members of Orion B9, might have been triggered by this feedback.} 



   \maketitle
%

\section{Introduction}

The protostellar phase of low-mass star formation begins when a starless 
(prestellar) core collapses, and, after a hypothesised short-lived 
first-hydrostatic core stage (\cite{larson1969}; \cite{masunaga1998}), 
a stellar embryo forms in its centre (the so-called second hydrostatic core; 
e.g., \cite{masunaga2000}). 
The dense cores harbouring the youngest protostars are known as the Class 0 
objects (\cite{andre1993}, 2000). In these objects, most of the system's mass 
resides in the dense envelope, i.e., $M_{\rm env}\gg M_{\star}$, where $M_{\star}$ 
is the mass of the central protostar. For this reason, Class 0 objects, or 
at least the youngest of them, are expected to still represent the initial 
physical conditions prevailing at the time of collapse phase.  
Class 0 objects are characterised by accretion-powered jets and mole\-cular 
outflows, which can be very powerful and highly collimated (e.g., 
\cite{bontemps1996}; \cite{gueth1999}; \cite{arce2005}, 2006; 
\cite{lee2007}). The statistical lifetime of the Class 0 stage is estimated to 
be $\sim1\times10^5$ yr (\cite{evans2009}; \cite{enoch2009}), but the exact 
duration of this embedded phase of evolution can be highly dependent on the 
initial/environmental conditions (e.g., \cite{vorobyov2010}).

\begin{figure}[!h]
\centering
\resizebox{0.8\hsize}{!}{\includegraphics{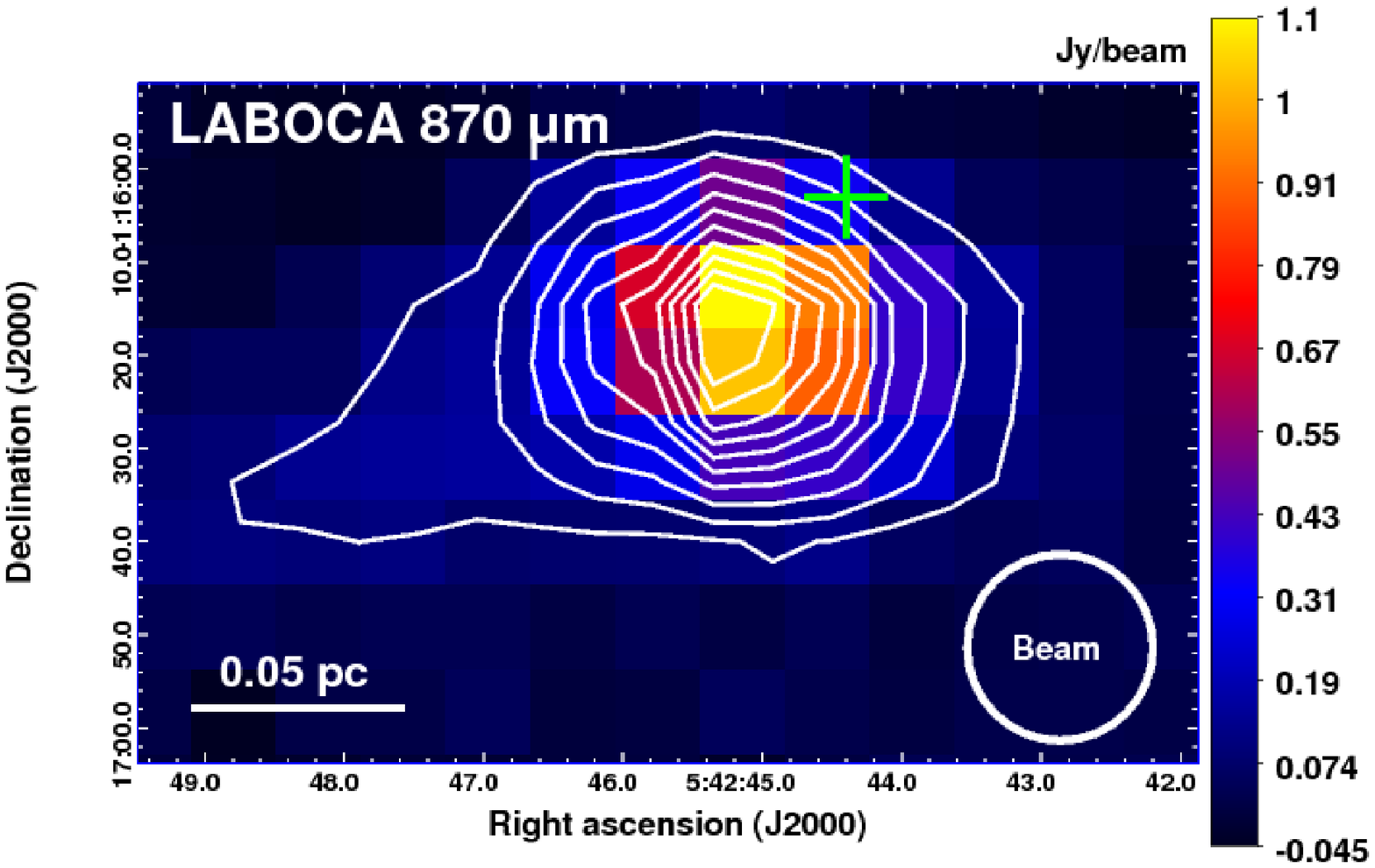}}
\resizebox{0.8\hsize}{!}{\includegraphics{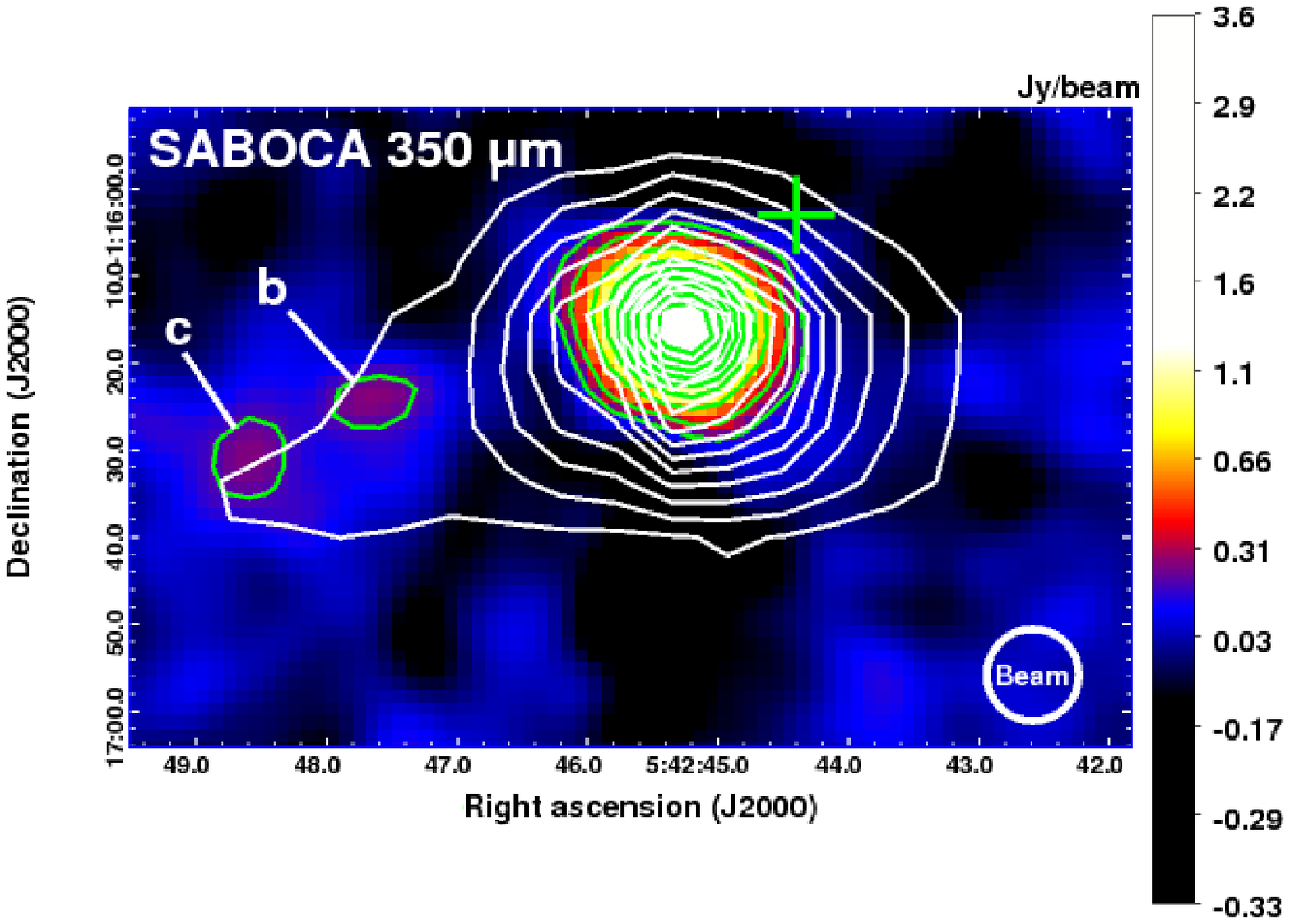}}
\resizebox{0.8\hsize}{!}{\includegraphics{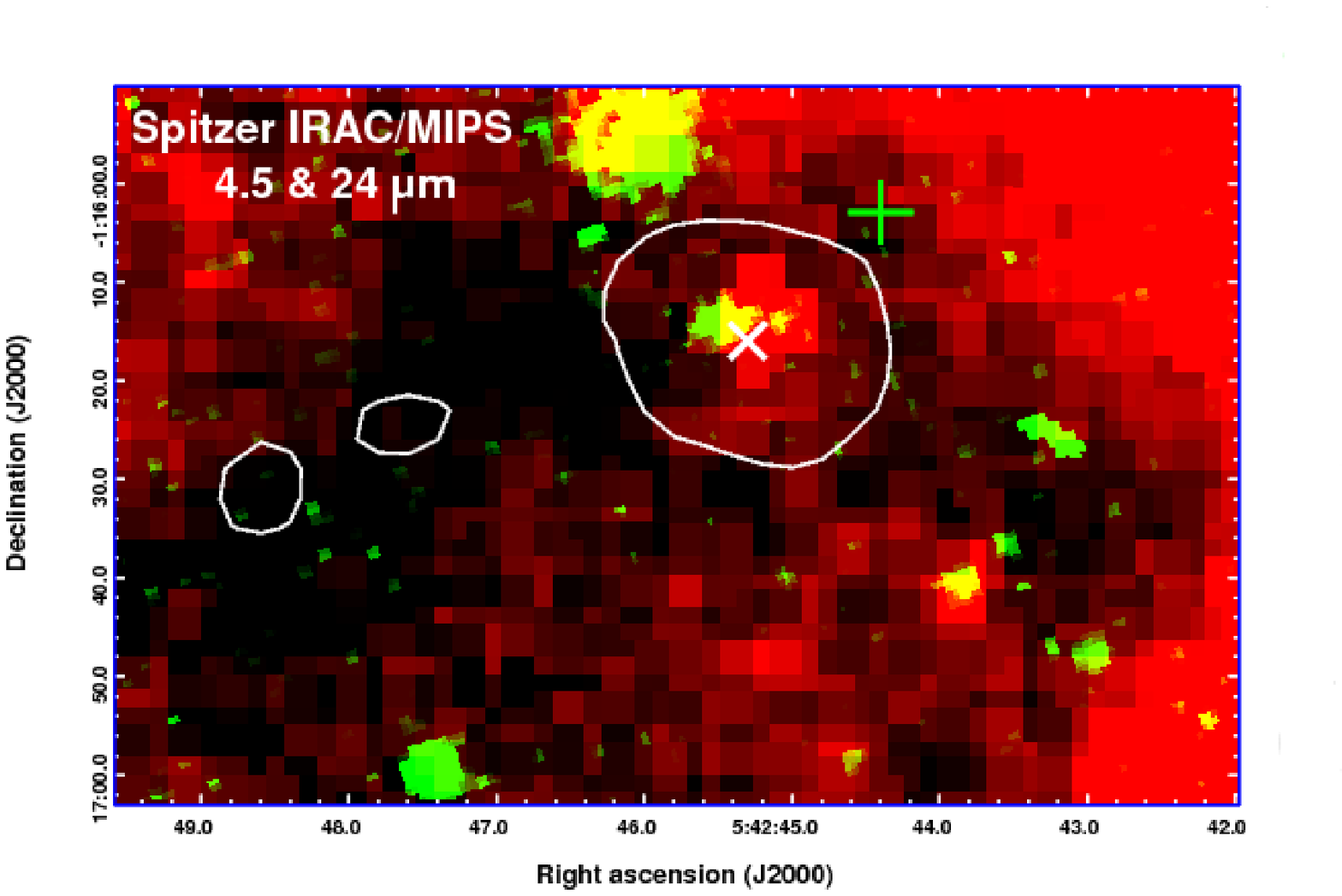}}
\caption{LABOCA 870-$\mu$m (\textit{top}), SABOCA 350-$\mu$m 
(\textit{middle}), and a \textit{Spitzer} IRAC/MIPS two-colour composite 
image (\textit{bottom}; 4.5 $\mu$m in green and 24 $\mu$m in red) of the 
Class  0 protostellar core SMM 3 in Orion B9. The LABOCA and \textit{Spitzer} 
images are shown with linear scaling, while the SABOCA image is shown with a 
square-root scaling to improve the contrast between bright and faint features. 
The LABOCA contours, plotted in white, go from 0.1 ($\sim3.3\sigma$) to 1.0 
Jy~beam$^{-1}$ in steps of 0.1 Jy~beam$^{-1}$. The SABOCA contour 
levels, plotted in green, start at $3\sigma$ and are 0.18 
Jy~beam$^{-1}\times [1,\,2,\,4,\,6,\,8,\,10,\,12,\,14,\,16]$. In the bottom 
panel, the first SABOCA contour, i.e., the $3\sigma$ emission level, is 
plotted in white to guide the eye, and the white cross indicates the SABOCA 
peak position of SMM 3. The small subcondensations SMM 3b and 3c discovered 
in Paper III are labeled in the middle panel. The green plus sign shows the 
target position of our previous molecular-line observations (i.e., the submm 
peak position of the LABOCA map before adjusting the pointing; see Paper III 
for details). A scale bar indicating the 0.05 pc projected length is shown in 
the bottom left of the top panel, with the assumption of a 450 pc 
line-of-sight distance. The effective LABOCA and SABOCA beams, 
$\sim20\arcsec$ and $10\farcs6$, are shown in the lower right corners of the 
corresponding panels.}
\label{figure:SMM3}
\end{figure}

The target source of the present study is the Class 0 protostellar core SMM 3 
in the Orion B9 star-forming region, which was originally discovered by 
Miettinen et al. (2009; Paper I) through LABOCA 870-$\mu$m dust continuum 
mapping of the region. SMM 3 is a strong submm emitting dust 
core ($S_{870}\simeq2.5$ Jy) that is associated with a weak \textit{Spitzer} 
24-$\mu$m point source ($S_{24}\simeq5$ mJy), and a 3.6 Jy 
point source at 70 $\mu$m. Using the Effelsberg 100-m telescope NH$_3$ 
observations, Miettinen et al. (2010; Paper II) derived the gas kinetic 
temperature of $T_{\rm kin}=11.3\pm0.8$ K in SMM 3. Using this temperature, 
the core mass was determined to be $7.8\pm1.6$ M$_{\sun}$, and its 
volume-averaged H$_2$ number density was estimated to be $1.1\pm0.2\times10^5$ 
cm$^{-3}$. In the SABOCA 350-$\mu$m mapping of Orion B9 by Miettinen et al. 
(2012; Paper III), SMM 3 was found to be by far the strongest source in the 
mapped area ($S_{350}\simeq5.4$ Jy). We also found that it contains two 
subfragments, or condensations (we called SMM 3b and 3c), lying about 
$36\arcsec-51\arcsec$ in projection from the central protostar. These 
correspond to 0.08--0.11 pc or $\sim1.7-2.3\times10^4$ AU at $d=450$ 
pc\footnote{In this paper, we adopt a distance of 450 pc to the Orion 
giant molecular cloud (\cite{genzel1989}). The actual distance may be somewhat 
smaller as, for example, Menten et al. (2007) determined a trigonometric 
parallax distance of $414\pm7$ pc to the Orion Nebula.}. 
Because the thermal Jeans length of the core is $\lambda_{\rm J}=0.07$ pc, we 
suggested that the core fragmentation into condensations can be explained by 
thermal Jeans instability. Using the 350/870-$\mu$m flux density ratio, 
we determined the dust temperature of the core to be 
$T_{\rm dust}=10.8_{-2.6}^{+5.7}$ K, which is very close to $T_{\rm kin}$ within the 
error bars. The revised spectral energy distribution (SED) of the core 
yielded a very low dust tempe\-rature of 8 K, and a bolometric luminosity 
of $L_{\rm bol}=1.2\pm0.1$ L$_{\sun}$. The latter is very close to the median 
luminosity of protostars in nearby star-forming regions, i.e., 
$L_{\rm med}=1.5^{+0.7}_{-0.4}$ L$_{\sun}$ (\cite{enoch2009}; \cite{offner2011}). 
In Paper III, we also studied the che\-mistry of SMM 3. We derived a large 
CO depletion factor of $f_{\rm D}({\rm CO})=10.8\pm2.2$, and a high level of 
deuterium fractionation, i.e., a ${\rm N_2D^+}/{\rm N_2H^+}$ column density 
ratio of $0.338\pm0.092$. In Fig.~\ref{figure:SMM3}, we show the LABOCA 
870-$\mu$m, SABOCA 350-$\mu$m, and \textit{Spitzer} 4.5/24-$\mu$m images 
towards SMM 3. In Table~\ref{table:SMM3}, we provide an overview of the 
physical and chemical properties of SMM 3 derived in our previous papers.

In this paper, we discuss the results of our $^{13}$CO and C$^{18}$O mapping 
observations of the environment of SMM 3. We ana\-lyse the structure of the 
mapped region as traced by emission from the $J=2-1$ rotational transition of 
the above CO isotopologues. The rest of the present paper is organised as 
follows. Observations and data reduction are described in Sect.~2. Mapping 
results and analysis are presented in Sect.~3. In Sect.~4, we discuss our 
results, and in Sect.~5, we summarise and conclude the paper.

\begin{table}
\renewcommand{\footnoterule}{}
\caption{Summary of the properties of SMM 3.}
\begin{minipage}{1\columnwidth}
\centering
\label{table:SMM3}
\begin{tabular}{c c}
\hline\hline 
Parameter & Value\\
\hline
{\bf SMM 3} \\
$\alpha_{2000.0}$\tablefootmark{a} & 05$^{\rm h}$ 42$^{\rm m}$ $45\fs3$ \\
$\delta_{2000.0}$\tablefootmark{a} & -01\degr 16\arcmin 16\arcsec \\
${\rm v}_{\rm LSR}$\tablefootmark{b} & $8.68\pm0.06$ km~s$^{-1}$\\
$R_{\rm eff}$\tablefootmark{c} & 13\farcs3 (0.03 pc)\\
$T_{\rm kin}$\tablefootmark{d} & $11.3\pm0.8$ K\\
$T_{\rm dust}$\tablefootmark{e} & $10.8_{-2.6}^{+5.7}$ K\\
$T_{\rm dust}^{\rm SED,\, cold}$ & 8.0 K\\
$\sigma_{\rm NT}$\tablefootmark{d} & $0.14\pm0.003$ km~s$^{-1}$\\
$\sigma_{\rm NT}/c_{\rm s}$\tablefootmark{d} & $0.7\pm0.04$ \\
$M$\tablefootmark{f} & $7.8\pm1.6$ M$_{\sun}$/$2.1\pm0.8$ M$_{\sun}$\\
$\alpha_{\rm vir}$\tablefootmark{g} & $0.5\pm0.1$\\
$N({\rm H_2})$\tablefootmark{f} & $8.4\pm1.1\times10^{22}$/$1.0\pm0.3\times10^{23}$ cm$^{-2}$\\
$\langle n({\rm H_2}) \rangle$\tablefootmark{f} & $1.1\pm0.2\times10^5$/$4.0\pm1.5\times10^5$ cm$^{-3}$\\
$L_{\rm bol}=L_{\rm cold}+L_{\rm warm}$ & $(0.3\pm0.1)+(0.9\pm0.1)=1.2\pm0.1$ L$_{\sun}$\\
$L_{\rm submm}/L_{\rm bol}$\tablefootmark{h} & 0.1 \\
$f_{\rm D}({\rm CO})$ & $10.8\pm2.2$ \\
$[{\rm N_2D^+}]/[{\rm N_2H^+}]$ & $0.338\pm0.092$ \\
{\bf SMM 3b} \\
$\alpha_{2000.0}$\tablefootmark{a} & 05$^{\rm h}$ 42$^{\rm m}$ $47\fs6$ \\
$\delta_{2000.0}$\tablefootmark{a} & -01\degr 16\arcmin 24\arcsec \\
$N({\rm H_2})$\tablefootmark{i} & $0.7\pm0.2\times10^{22}$ cm$^{-2}$\\
{\bf SMM 3c} \\
$\alpha_{2000.0}$\tablefootmark{a} & 05$^{\rm h}$ 42$^{\rm m}$ $48\fs6$ \\
$\delta_{2000.0}$\tablefootmark{a} & -01\degr 16\arcmin 32\arcsec \\
$N({\rm H_2})$\tablefootmark{i} & $0.7\pm0.2\times10^{22}$ cm$^{-2}$\\
\hline 
\end{tabular} 
\tablefoot{\tablefoottext{a}{SABOCA 350-$\mu$m peak 
position.}\tablefoottext{b}{The LSR velocity derived from optically thin 
C$^{17}$O$(2-1)$ line.}\tablefoottext{c}{Effective radius of the ``main'' core 
as determined from the SABOCA 350-$\mu$m map.}\tablefoottext{d}{Derived from 
NH$_3$ data. $\sigma_{\rm NT}$ and $c_{\rm s}$ are, respectively, the one 
dimensional non-thermal velocity dispersion and the isothermal sound 
speed.}\tablefoottext{e}{Computed from the 350-to-870 $\mu$m flux density 
ratio.}\tablefoottext{f}{The first value refers to the LABOCA 870-$\mu$m 
core, and the second one to the ``main'' core detected at 350 
$\mu$m.}\tablefoottext{g}{Virial parameter defined by 
$\alpha_{\rm vir}=M_{\rm vir}/M$.}\tablefoottext{h}{$L_{\rm submm}$ is the submm 
luminosity derived by integrating the SED longward of 
350 $\mu$m.}\tablefoottext{i}{Calculated by making the assumption that 
$T_{\rm dust}$ equals the $T_{\rm kin}$ derived for the ``main'' core.}}
\end{minipage} 
\end{table}

\section{Observations and data reduction}

The observations presented in this paper were made on 13 November 2011 using 
the APEX 12-m telescope located at Llano de Chajnantor in the Atacama desert 
of Chile. The telescope and its performance are described in the paper by 
G{\"u}sten et al. (2006). An area of $4\arcmin \times 4\arcmin$ (0.52 pc 
$\times$ 0.52 pc at $d=450$ pc) was simultaneously 
mapped in the $J=2-1$ rotational lines of $^{13}$CO and C$^{18}$O using the 
total power on-the-fly mode towards SMM 3 centred on the coordinates 
$\alpha_{2000.0}=05^{\rm h}42^{\rm m}45\fs8$, 
$\delta_{2000.0}=-01\degr 16\arcmin 13\farcs0$ [$(+7\farcs5,\,+3\farcs0)$ 
offset from the SABOCA peak position of SMM3]. At the $^{13}$CO$(2-1)$ and 
C$^{18}$O$(2-1)$ line frequencies, 220\,398.70056 and 
219\,560.357 MHz\footnote{The $^{13}$CO$(2-1)$ frequency was taken from 
Cazzoli et al. (2004), and it refers to the strongest hyperfine component 
$F=5/2-3/2$. The C$^{18}$O$(2-1)$ frequency was adopted from the JPL 
spectroscopic database at {\tt http://spec.jpl.nasa.gov/} 
(\cite{pickett1998}).}, respectively, the telescope beam size is about 
$28\farcs3$ (HPBW). The target area was scanned alternately in right ascension 
and declination, i.e., in zigzags to ensure minimal striping artefacts in the 
final data cubes. Both the stepsize between the subscans and the angular 
separation between two successive dumps was $9\farcs4$, i.e., about 1/3 times 
the beam HPBW ensuring Nyquist sampling. We note that the readout spacing 
$1/3 \times {\rm HPBW}$ should not be exceeded to avoid beam smearing. The 
integration time per dump and per pixel was 1 s.

As a frontend, we used the APEX-1 receiver of the 
Swedish Heterodyne Facility Instrument (SHeFI; \cite{belitsky2007}; 
\cite{vassilev2008a},b). The backend was the RPG eXtended bandwidth Fast 
Fourier Transfrom Spectrometer (XFFTS; cf. \cite{klein2012}) with an 
instantaneous bandwidth of 2.5 GHz and 32\,768 spectral channels.
The resulting channel separation, 76.3 kHz, corresponds to about 0.1 
km~s$^{-1}$ at 220 GHz.

The telescope pointing accuracy was checked by CO$(2-1)$ cross maps of the 
variable star RAFGL865 (V1259 Ori), and was found to be consistent within 
$\lesssim4\arcsec$. The focus was checked by measurements on Jupiter. 
Calibration was made by means of the chopper-wheel technique and the output 
intensity scale given by the system is $T_{\rm A}^{\star}$, which represents the 
antenna temperature corrected for the atmospheric attenuation. The amount of 
precipitable water vapour (PWV) was in the range 1.28 -- 1.48 mm, and the 
single-sideband system temperature was around 150 K (in $T_{\rm A}^{\star}$ 
units). The observed intensities were converted to the main-beam brightness 
temperature scale by $T_{\rm MB}=T_{\rm A}^{\star}/\eta_{\rm MB}$, where 
$\eta_{\rm MB}=0.75$ is the main-beam efficiency at the frequencies used. The 
absolute calibration uncertainty is estimated to be about 10\%.

The spectra were reduced and the maps were produced using the CLASS90 and GREG 
programmes of the GILDAS software package\footnote{Grenoble Image and 
Line Data Analysis Software is provided and actively developed by IRAM, and 
is available at {\tt http://www.iram.fr/IRAMFR/GILDAS}}. 
The individual spectra were Hanning-smoothed to improve the 
signal-to-noise ratio of the data. A third-order polynomial was applied to 
correct the baseline in the spectra. The resulting $1\sigma$ rms noise level 
of the average spectra are about 30 mK (in $T_{\rm MB}$) at the smoothed 
resolution (16\,384 channels). The data were convolved with a Gaussian of 
1/3 times the beam HPBW, and therefore the effective angular resolution 
of the final data cubes is about $30\arcsec$. The average $1\sigma$ rms noise 
level of the completed maps ranges from 0.20 to 0.28 K per 0.2 km~s$^{-1}$ 
channel on a $T_{\rm MB}$ scale. The constructed data cubes were exported in 
FITS format for further processing in IDL.

\section{Mapping results and analysis}

\subsection{The average spectra}

\begin{figure}[!h]
\centering
\resizebox{0.75\hsize}{!}{\includegraphics{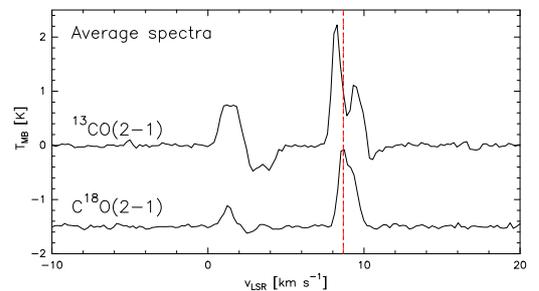}}
\caption{Hanning-smoothed spatially averaged $^{13}$CO$(2-1)$ and 
C$^{18}$O$(2-1)$ spectra across the mapped field. The C$^{18}$O$(2-1)$ spectrum 
is offset by -1.45 K from zero baseline for reasons of clarity. The vertical 
red dashed line indicates the systemic velocity of SMM 3 derived from 
C$^{17}$O$(2-1)$ in Paper III.}
\label{figure:average}
\end{figure}

The average $^{13}$CO$(2-1)$ and C$^{18}$O$(2-1)$ spectra are shown in 
Fig.~\ref{figure:average}. Both lines exhibit two well-separated velocity 
components: one near the systemic velocity of about 8.7 km~s$^{-1}$, and 
the other at 1.3--1.4 km~s$^{-1}$. It is not surprising that we see these 
lower-velocity components in the lines of CO isotopologues. The additional 
lines at comparable radial velocities of 1.3--1.9 km~s$^{-1}$ were already 
detected in the lines of N$_2$H$^+(1-0)$ and N$_2$D$^+(2-1)$ in Paper I, 
NH$_3(1,\,1)$ and $(2,\,2)$ in Paper II, and C$^{17}$O$(2-1)$, 
H$^{13}$CO$^+(4-3)$, DCO$^+(4-3)$, N$_2$H$^+(3-2)$, and N$_2$D$^+(3-2)$ in 
Paper III towards other cores in Orion B9. Therefore, detection of 
lower-velocity line emission from CO isotopologues was expected. 
We note that the average $^{13}$CO line near the systemic velocity of SMM 3 
appears to show a blue asymmetric profile with blue peak being stronger than 
the red peak. The central dip also appears to be near the radial velocity 
derived from optically thin C$^{17}$O$(2-1)$ line in Paper III. Despite of 
these characteristics, the double-peaked line profile is \textit{not} caused by 
infall motions (e.g., \cite{myers1996}); it results from averaging over the 
entire mapped area, where two separate velocity components at about 8.5 and 
9.5 km~s$^{-1}$ are seen. This issue will be further discussed in Sect.~3.3. 
The average C$^{18}$O line profile at the systemic 
velocity is nearly Gaussian, which suggests that the line is likely to be 
optically thin. However, a hint of the two nearby velocity components 
is also visible in the average C$^{18}$O line; the line exhibits a small 
``knee'' at $\sim9.5$ km~s$^{-1}$.

It can also be seen from Fig.~\ref{figure:average} that some of the 
observation OFF positions had $^{13}$CO emission in the velocity regime 
bet\-ween about 2.4 and 4.8 km~s$^{-1}$, and between about 10.3 and 11.3 
km~s$^{-1}$, and C$^{18}$O emission in the range 2.0--3.3 km~s$^{-1}$. These 
velocity regimes show up as artificial absorption features in the average 
spectra. Given the ubiquitous nature of multiple velo\-city components along 
the line of sight towards Orion B9, finding an emission free OFF position from 
this region can be difficult.
 
The main purpose of examining the average spectra is to determine the velocity 
range of the detected emission. This is needed to construct the line-emission 
maps, as will be described in the next section. 

\subsection{Moment maps}

To display the intensity and kinematic structure of the $^{13}$CO and C$^{18}$O 
line emission, we constructed the moment maps by integrating the lines over 
the following LSR velocity ranges: $[6.09,\,10.36]$ km~s$^{-1}$ and 
$[7.77,\,10.36]$ km~s$^{-1}$ for the $^{13}$CO and C$^{18}$O lines of the main 
velocity component, and $[-0.07,\,2.52]$ km~s$^{-1}$ and $[0.14,\,2.10]$ 
km~s$^{-1}$ for the $^{13}$CO and C$^{18}$O lines of the lower-velocity 
component (at $\sim1.3$ km~s$^{-1}$). These line windows were selected so that 
the artificial absorption features discussed above are avoided. The threshold 
used for the moment maps was chosen to be 2 times the rms noise, i.e., 
$2\sigma$.

\begin{figure*}
\begin{center}
\includegraphics[width=0.33\textwidth]{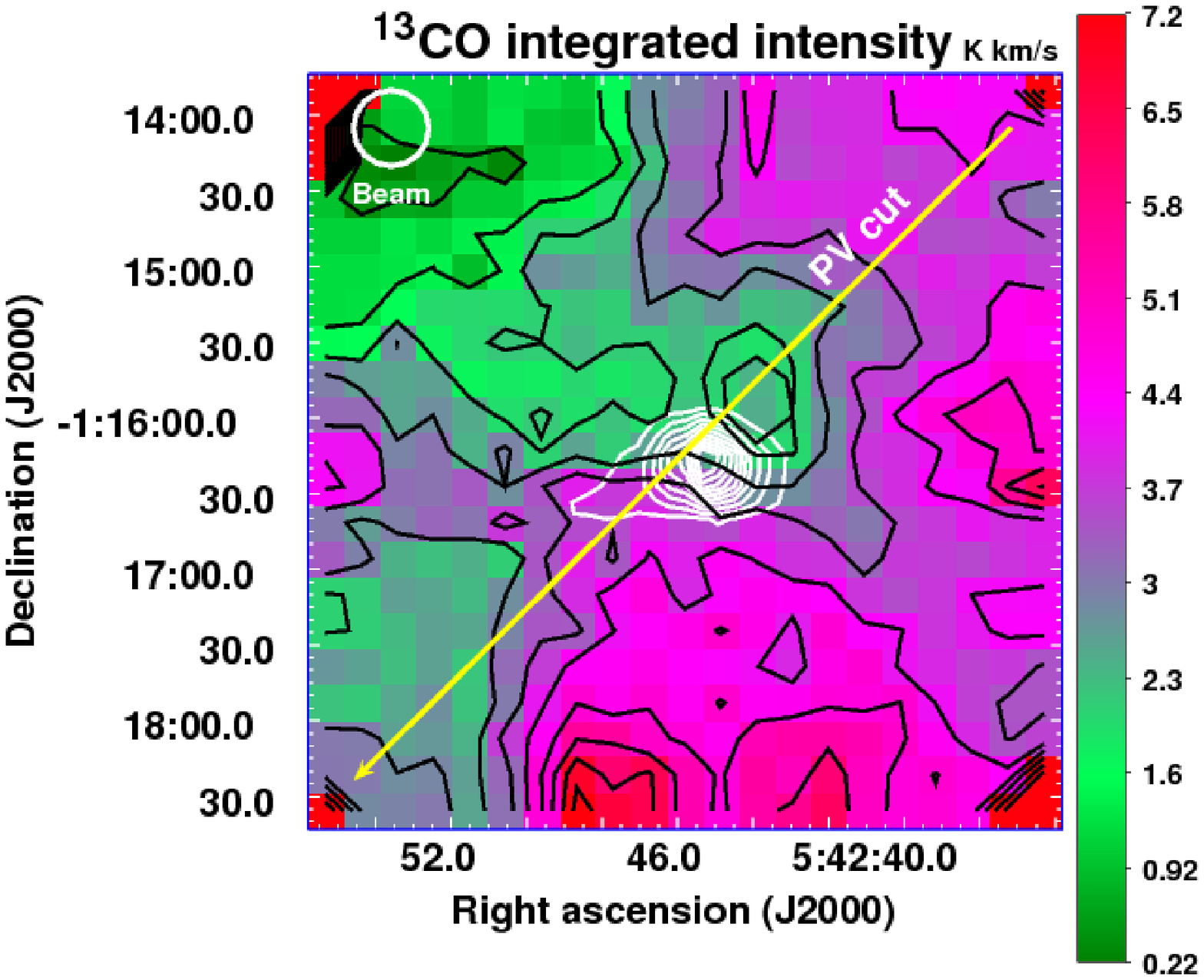}
\includegraphics[width=0.33\textwidth]{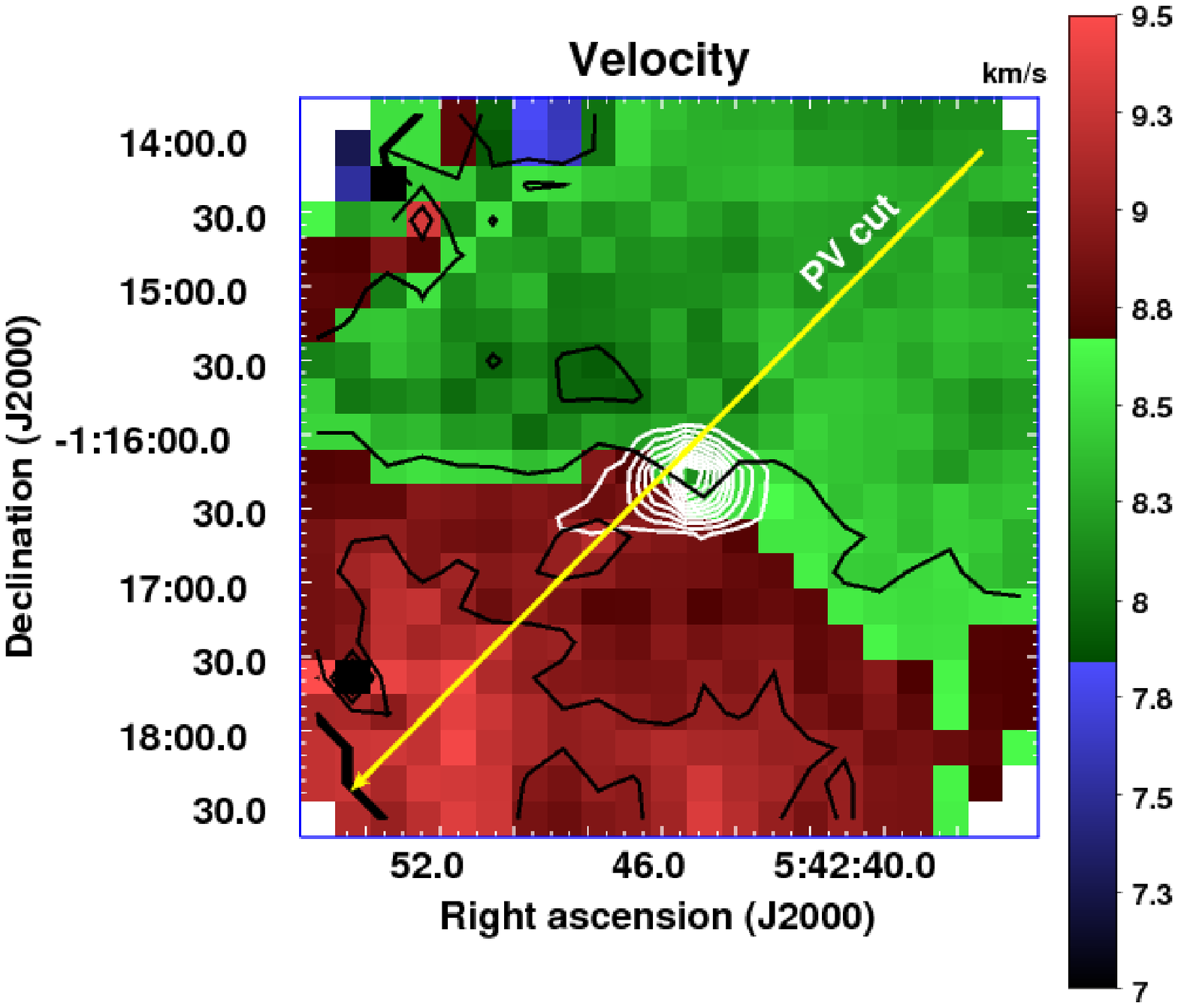}
\includegraphics[width=0.33\textwidth]{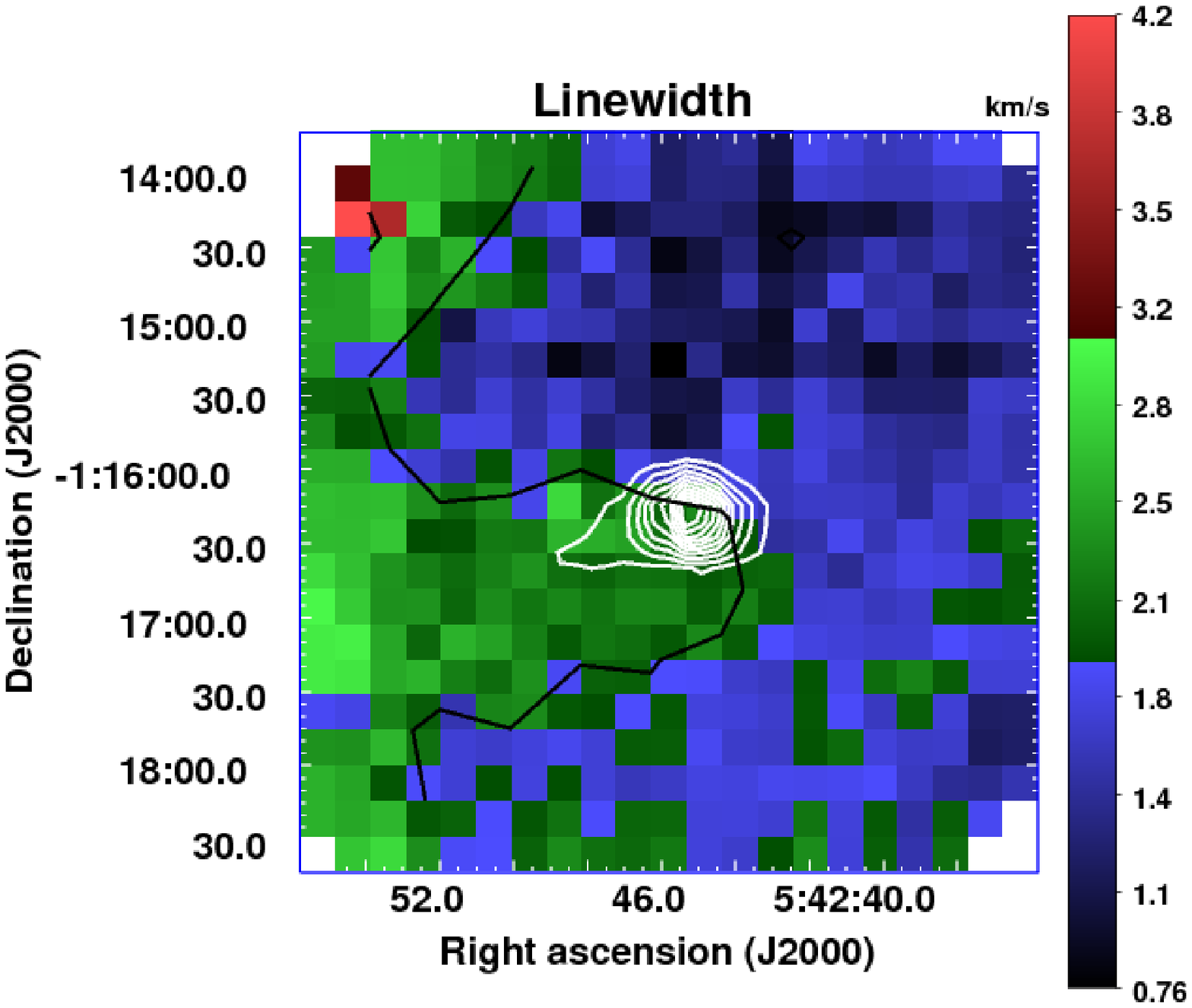}
\includegraphics[width=0.33\textwidth]{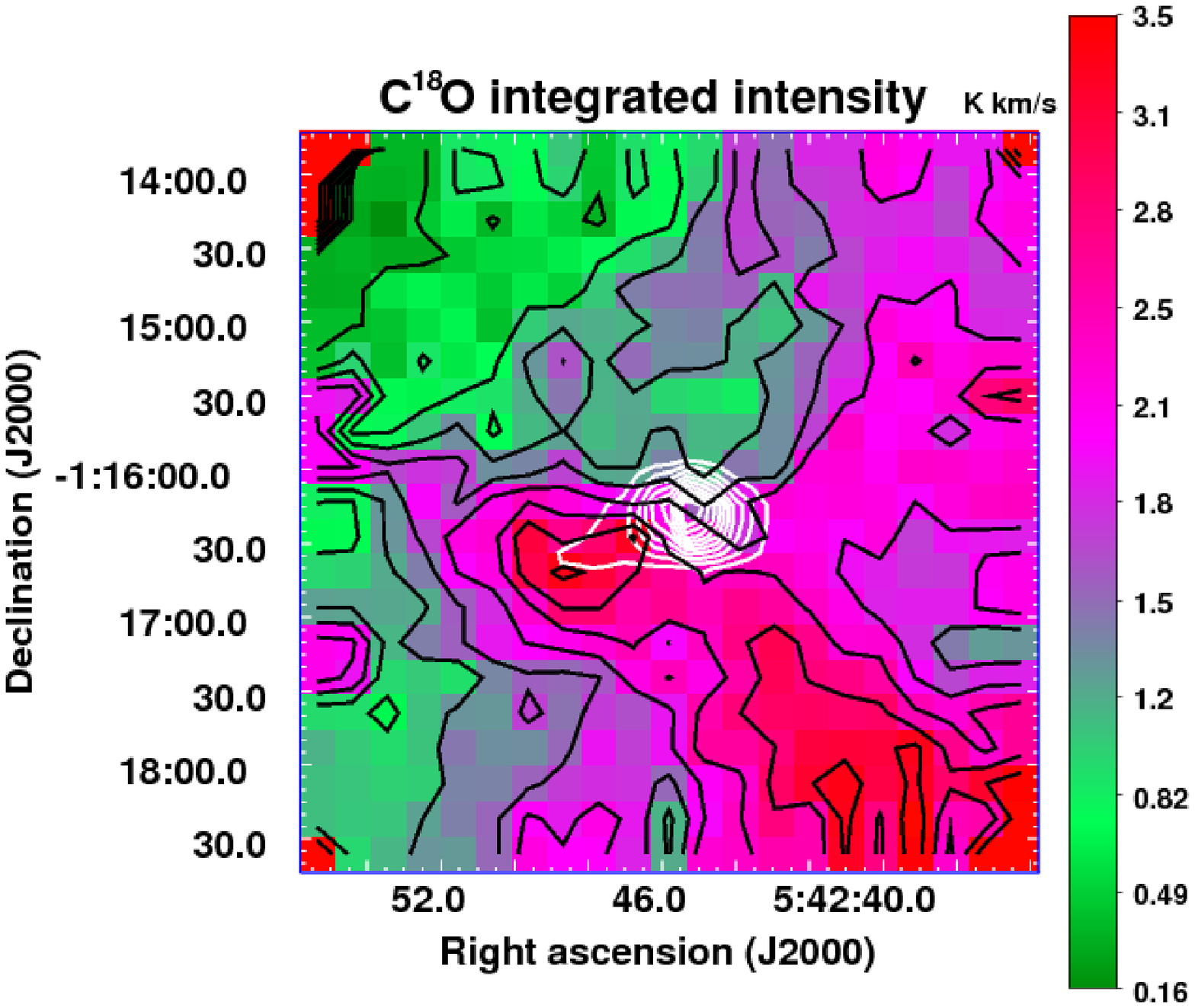}
\includegraphics[width=0.33\textwidth]{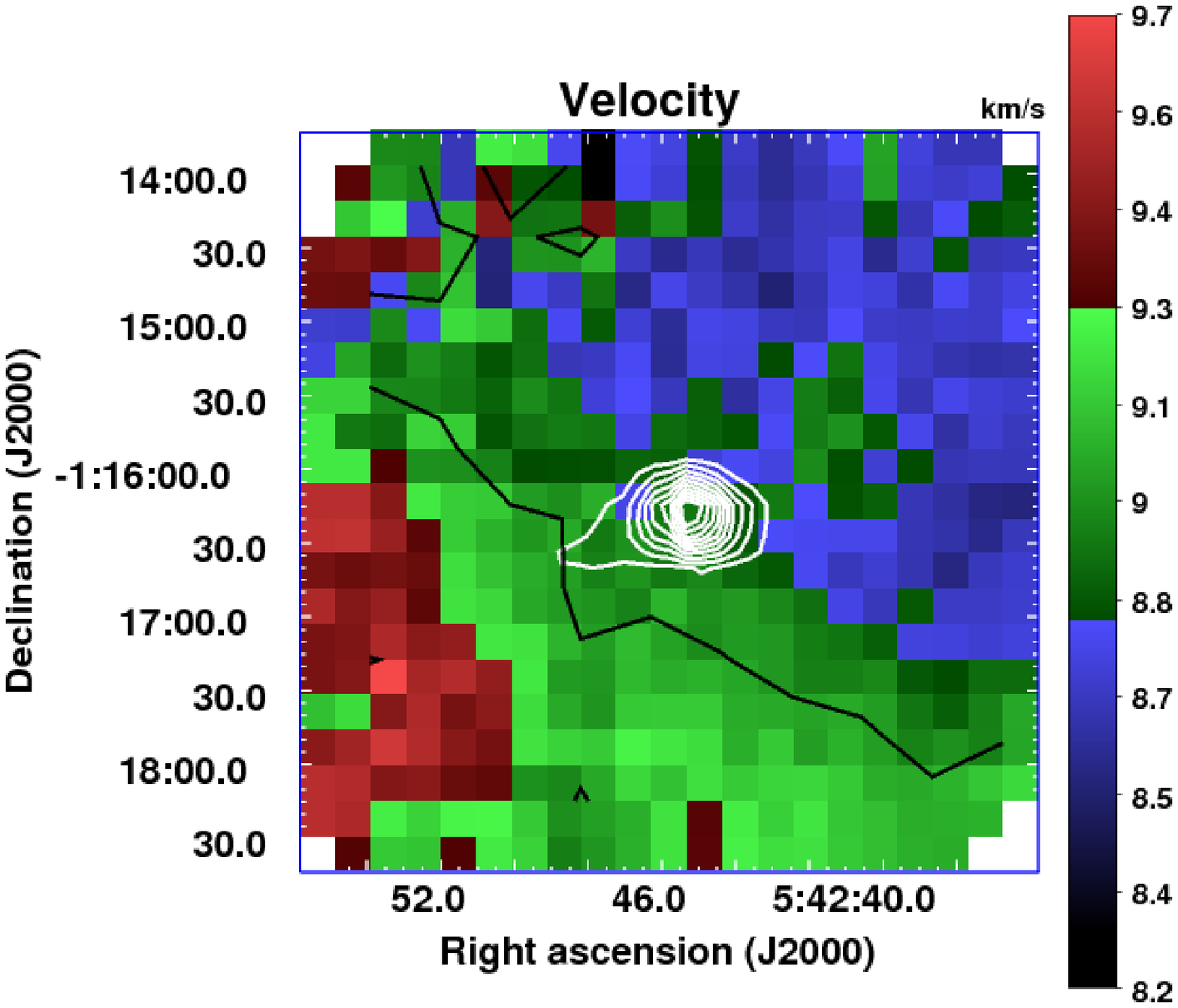}
\includegraphics[width=0.33\textwidth]{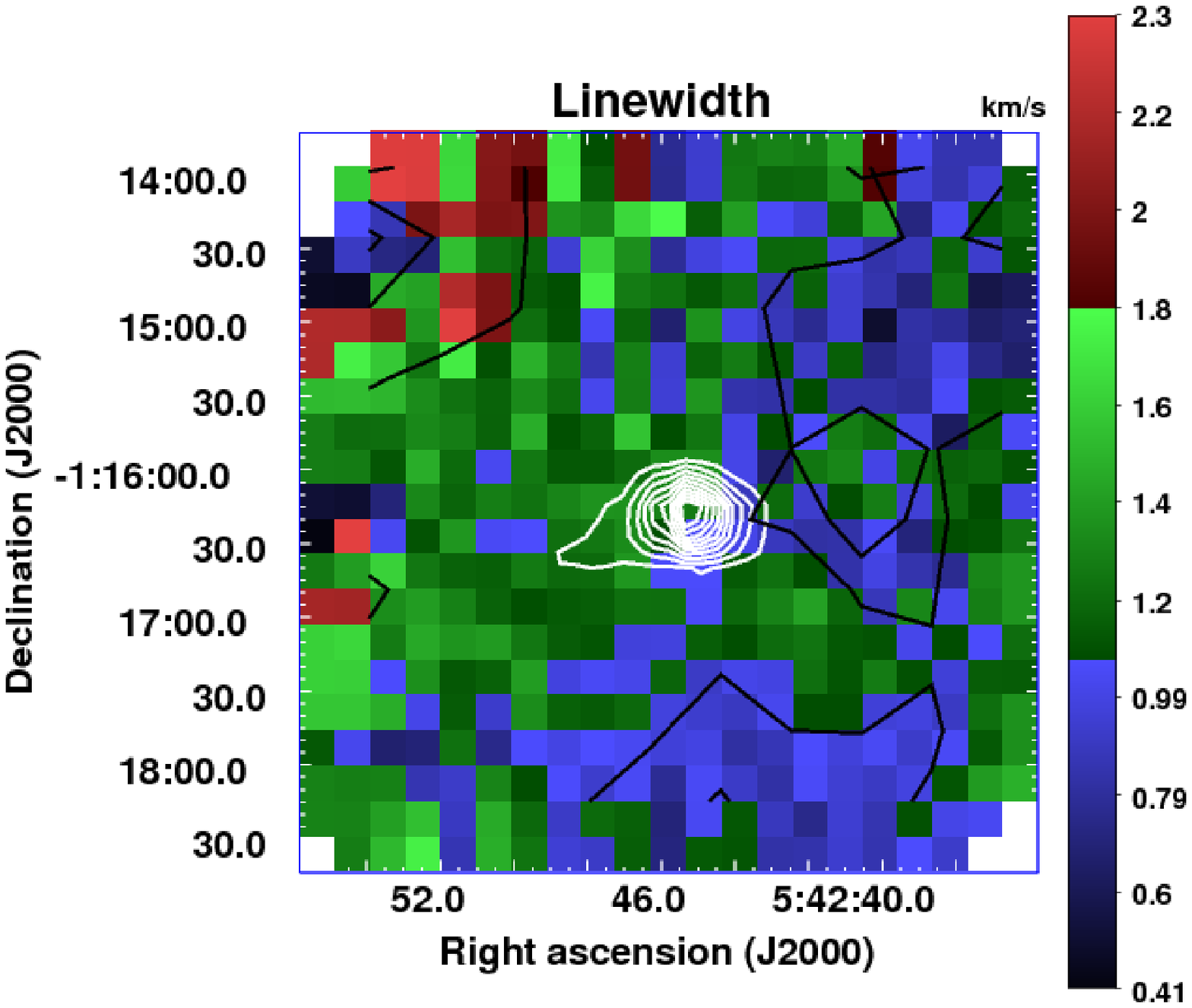}
\caption{The colour scale in the panels from left to right shows, 
respectively, the 0th (integrated intensity in $T_{\rm MB}$), 1st 
(intensity-weighted centroid velocity structure), and 2nd moment 
(intensity-weighted linewidth structure) map of the 8.7-km~s$^{-1}$ 
component $^{13}$CO$(2-1)$ (\textit{top row}) and C$^{18}$O$(2-1)$ emission 
(\textit{bottom row}). The black contours in the upper panels, from left to 
right, go as follows: from 0.7 ($3.2\sigma$) to 7.0 K~km~s$^{-1}$ ($32\sigma$) 
in steps of 0.7 K~km~s$^{-1}$; from 7 to 9.5 km~s$^{-1}$ in steps of 0.5 
km~s$^{-1}$; and from 1 to 4 km~s$^{-1}$ in steps of 1 km~s$^{-1}$. The 
corresponding contours in the lower panels are from 0.4 ($2\sigma$) to 
3.2 K~km~s$^{-1}$ ($16\sigma$) in steps of 0.4 K~km~s$^{-1}$; from 7 to 
9.5 km~s$^{-1}$ in steps of 0.5 km~s$^{-1}$; and from 0.5 to 2 km~s$^{-1}$ in 
steps of 0.5 km~s$^{-1}$. Superimposed on the maps are the LABOCA 870-$\mu$m 
contours in white as in Fig.~\ref{figure:SMM3}. The yellow solid line 
on the top left and middle panels shows the location of the PV slice extracted 
for Fig.~\ref{figure:pv}. The effective beam size of $30\arcsec$ is shown in 
the upper left corner of the upper left panel.}
\label{figure:moments}
\end{center}
\end{figure*}

\begin{figure*}
\begin{center}
\includegraphics[width=0.33\textwidth]{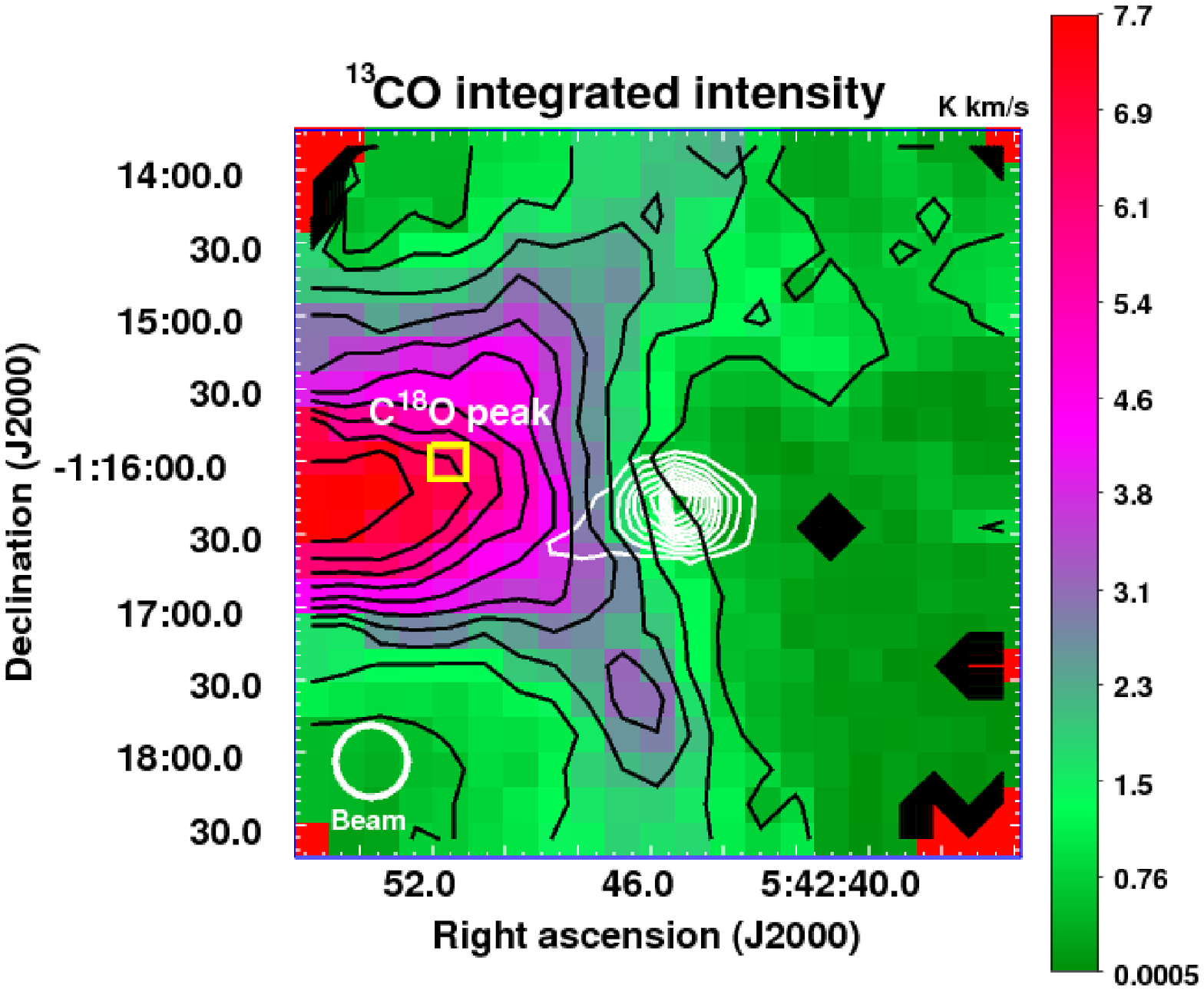}
\includegraphics[width=0.33\textwidth]{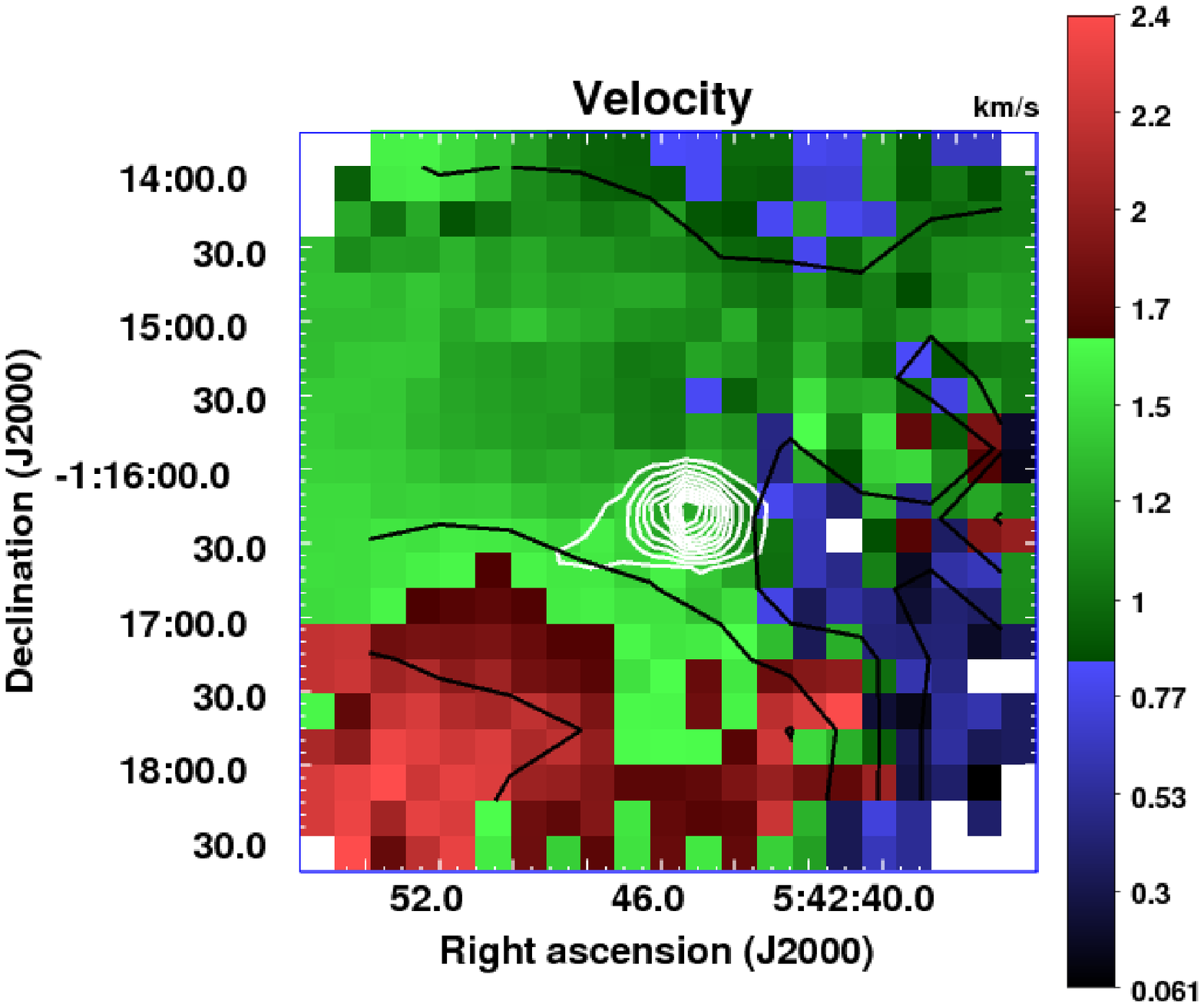}
\includegraphics[width=0.33\textwidth]{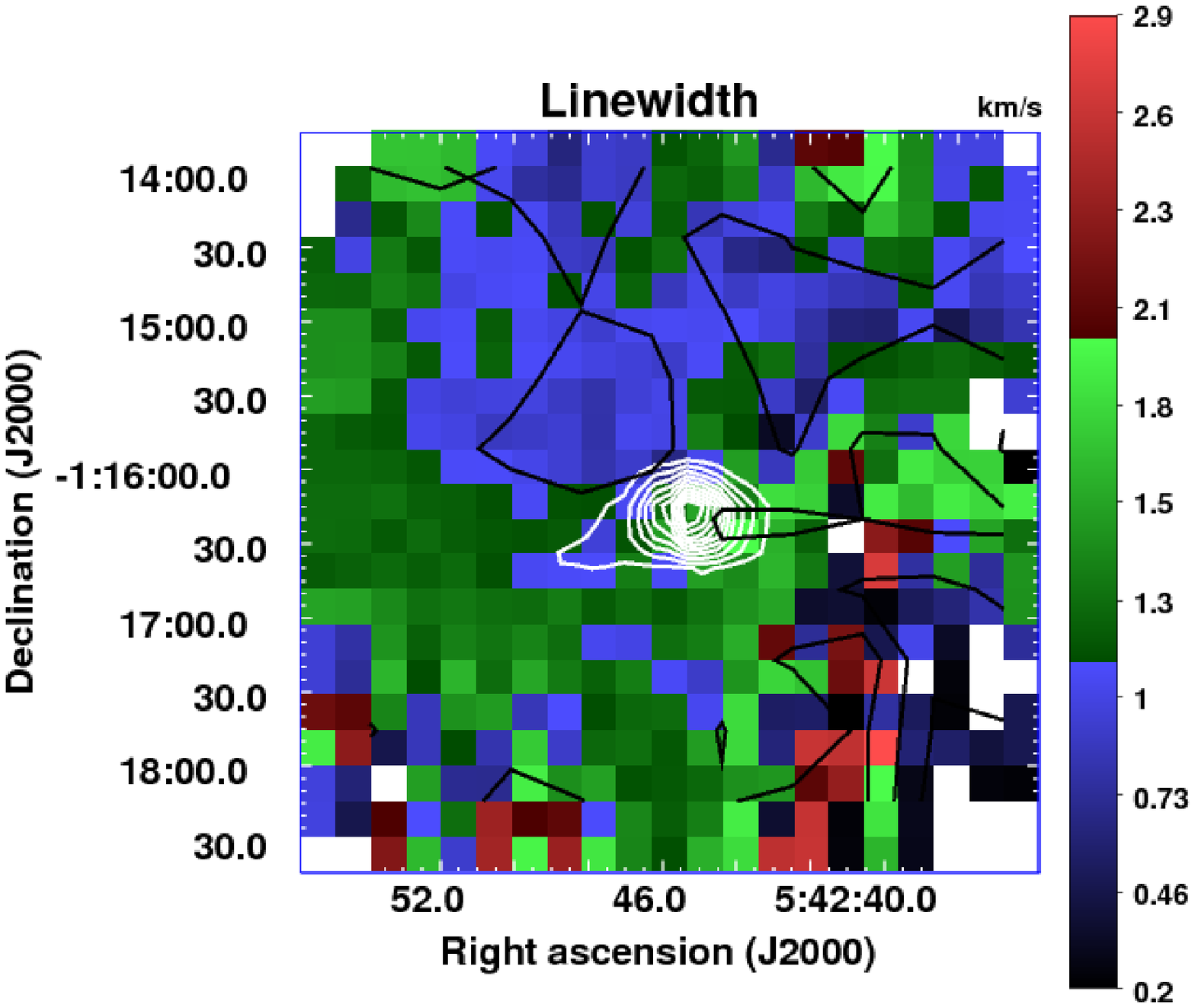}
\includegraphics[width=0.33\textwidth]{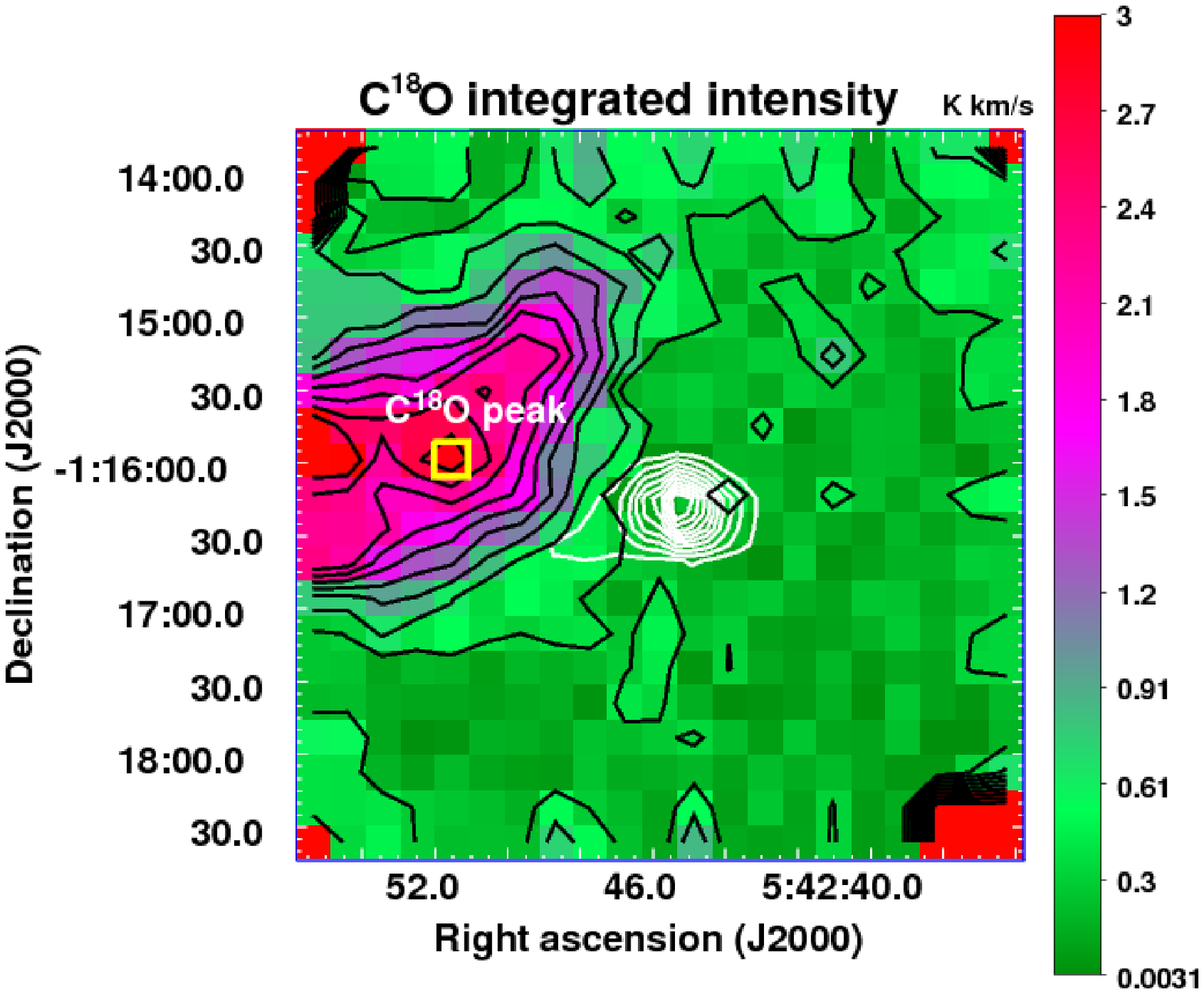}
\includegraphics[width=0.33\textwidth]{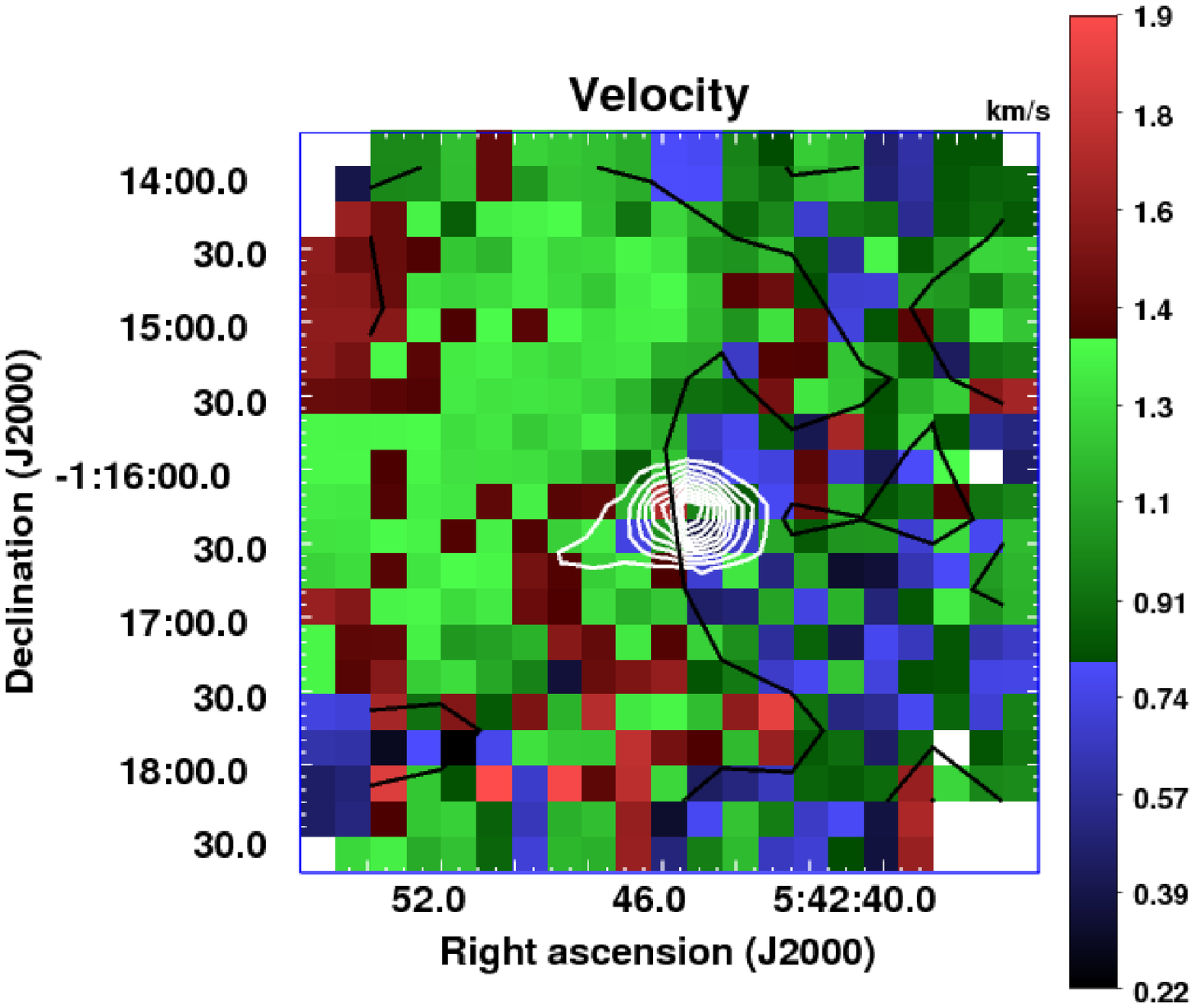}
\includegraphics[width=0.33\textwidth]{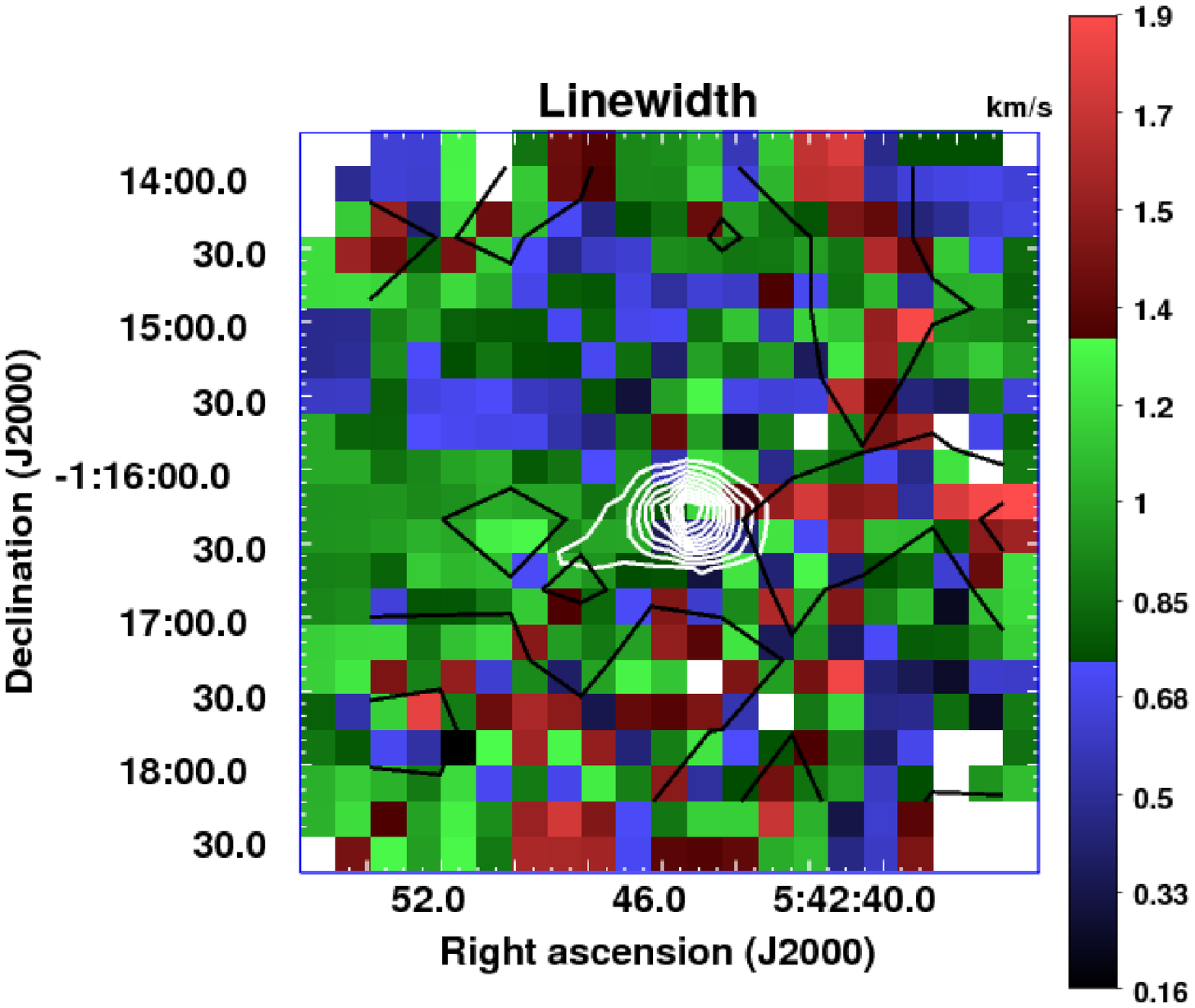}
\caption{Same as Fig.~\ref{figure:moments} but for the 1.3-km~s$^{-1}$ 
component. The black contours in the upper panels, from left to right, are 
plotted as follows: from 0.7 ($2.7\sigma$) to 7.0 K~km~s$^{-1}$ 
($27\sigma$) in steps of 0.7 K~km~s$^{-1}$; at 0.5, 1, 1.5, and 
2 km~s$^{-1}$; and from 0.5 to 2.5 km~s$^{-1}$ in steps of 0.5 km~s$^{-1}$. 
In the lower panels, these contours are from 0.3 ($1.1\sigma$) to 
2.7 K~km~s$^{-1}$ ($9.6\sigma$) in steps of 0.3 K~km~s$^{-1}$; 
at 0.5, 1, 1.5, and 2 km~s$^{-1}$; and at 0.5, 1, and 1.5 km~s$^{-1}$. 
The C$^{18}$O peak position, from which the 1.3-km~s$^{-1}$ component 
spectra were extracted (Sect.~3.4), is indicated by a yellow box in the left 
panels. The effective beam size $30\arcsec$ is shown in the lower left corner 
of the upper left panel.}
\label{figure:moments_second}
\end{center}
\end{figure*}

The zeroth, first, and second moment maps (i.e., the images of 
integrated intensity, intensity-weighted central velo\-city, and 
intensity-weighted FWHM linewidth) of the $^{13}$CO and C$^{18}$O emission of 
the main velocity component are shown in Fig.~\ref{figure:moments}, while 
those of the lower-velocity component are shown in 
Fig.~\ref{figure:moments_second}.

The map of $^{13}$CO integrated intensity (upper left panel of 
Fig.~\ref{figure:moments}) shows that there is a local emission minimum close 
to the SMM 3 protostellar position. This conforms to the high level of CO 
depletion derived from C$^{17}$O data. The $^{13}$CO emission around SMM 3 is 
rather extended, which is not surprising because it arises from the 
lower-density gas. In contrast, the LABOCA dust continuum emission shows only 
the densest part of the region, i.e., SMM 3. The $^{13}$CO emission appears to 
be strongest at about $2\arcmin$ south of SMM 3, and from there it extends to 
the northwest part of the map. From the lower left panel of 
Fig.~\ref{figure:moments}, it can be seen that the C$^{18}$O emission follows 
quite well the morphology of the $^{13}$CO emission. There is a hint of an 
elongated filament-type feature along the NE to the SW direction with 
relatively strong emission. There appears to be a few C$^{18}$O maxima at the 
eastern part of SMM 3, one lying at the eastern tip of the $3.3\sigma$ LABOCA 
contour, and the other about $24\arcsec$ from the central protostar.

In the case of the lower-velocity component, both the $^{13}$CO and 
C$^{18}$O emission are less extended, and are instead concentrated into a 
single clump-like feature at the eastern part of the mapped region (left panels 
of Fig.~\ref{figure:moments_second}). There is a $^{13}$CO extension 
to the south of SMM 3 in projection, which is not seen in C$^{18}$O.

Interestingly, as can be seen from the first-order moment map of 
$^{13}$CO (top middle panel of Fig.~\ref{figure:moments}), there 
is a fairly sharp border between the two velocity fields, and SMM 3 appears to 
lie exactly between them, i.e., at the border of the velocity gradient. There 
is also a hint of increasing $^{13}$CO linewidth across this border as shown 
in the top right panel of Fig.~\ref{figure:moments}. The radial-velocity 
structure of C$^{18}$O emission is quite similar to that of $^{13}$CO, but the 
C$^{18}$O linewidths show less obvious spatial trend (lower middle and 
right panels of Fig.~\ref{figure:moments}).

Another interesting feature concerning the radial-velocity distribution is 
that also the lower-velocity component of C$^{18}$O shows a somewhat similar 
gradient across the map (Fig.~\ref{figure:moments_second}; lower middle 
panel). Implications of the velocity gradient will be discussed further in 
Sect.~4.3.




\subsection{Channel maps and PV plots}

\begin{figure}[!h]
\centering
\resizebox{1.0\hsize}{!}{\includegraphics{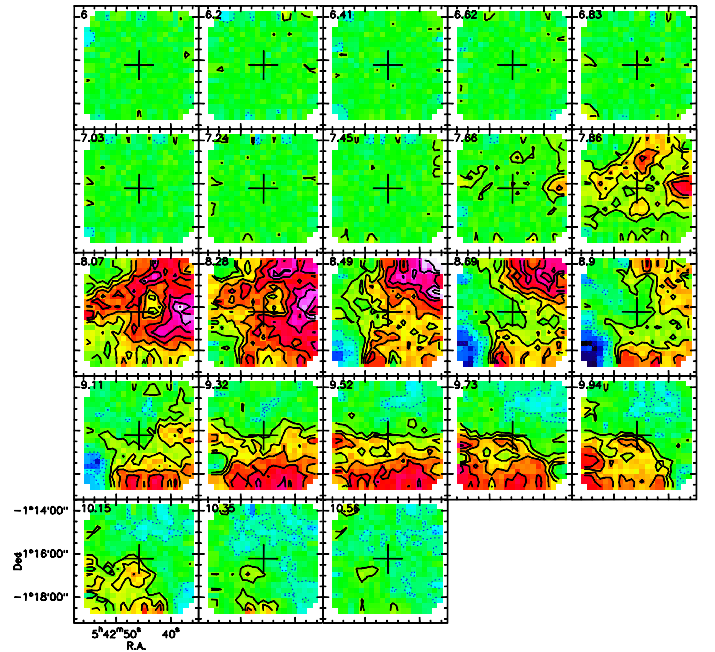}}
\caption{Velocity channel map across 6--10.56 km~s$^{-1}$ of the 
$^{13}$CO$(2-1)$ line towards SMM 3. The velocity of each channel is shown in 
the top-left corner of each panel. The solid contours go from 0.67 to 4.67 K in 
steps of 0.67 K ($T_{\rm MB}$), while the dashed contours mirror negative 
values due to the baseline-fitting problems. The central cross on 
each panel indicates the map centre.}
\label{figure:channel1}
\end{figure}

\begin{figure}[!h]
\centering
\resizebox{1.0\hsize}{!}{\includegraphics{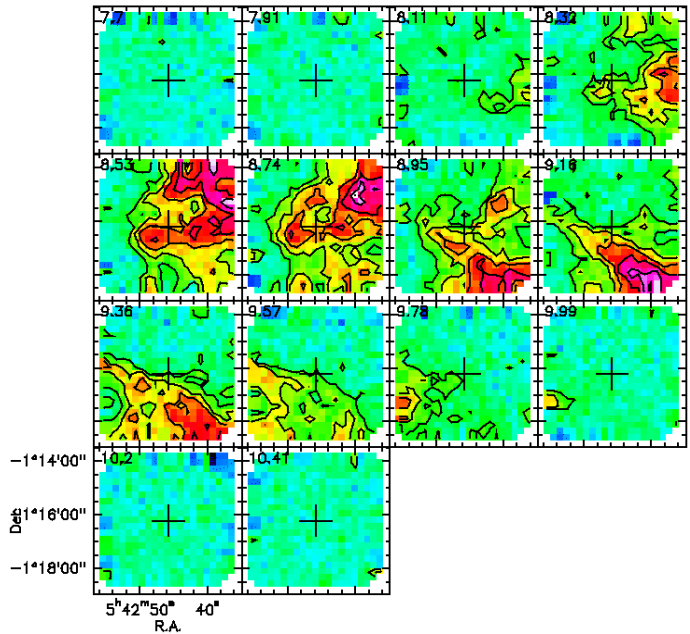}}
\caption{Same as Fig.~\ref{figure:channel1} but across 7.7--10.41 km~s$^{-1}$ 
of the C$^{18}$O$(2-1)$ line. The contours are as in 
Fig.~\ref{figure:channel1}.}
\label{figure:channel2}
\end{figure}

Velocity channel maps of the $^{13}$CO and C$^{18}$O emission for the main 
velocity component are plotted in Figs.~\ref{figure:channel1} and 
\ref{figure:channel2}, respectively. As shown in these maps, extended 
$^{13}$CO emission can be seen across the velocity range 
$7.86 < {\rm v}_{\rm LSR} < 10.15$ km~s$^{-1}$, while that of C$^{18}$O extends 
over a somewhat narrower range of velocities, $8.32 < {\rm v}_{\rm LSR} < 9.78$ 
km~s$^{-1}$. 

\begin{figure}[!h]
\centering
\resizebox{0.8\hsize}{!}{\includegraphics{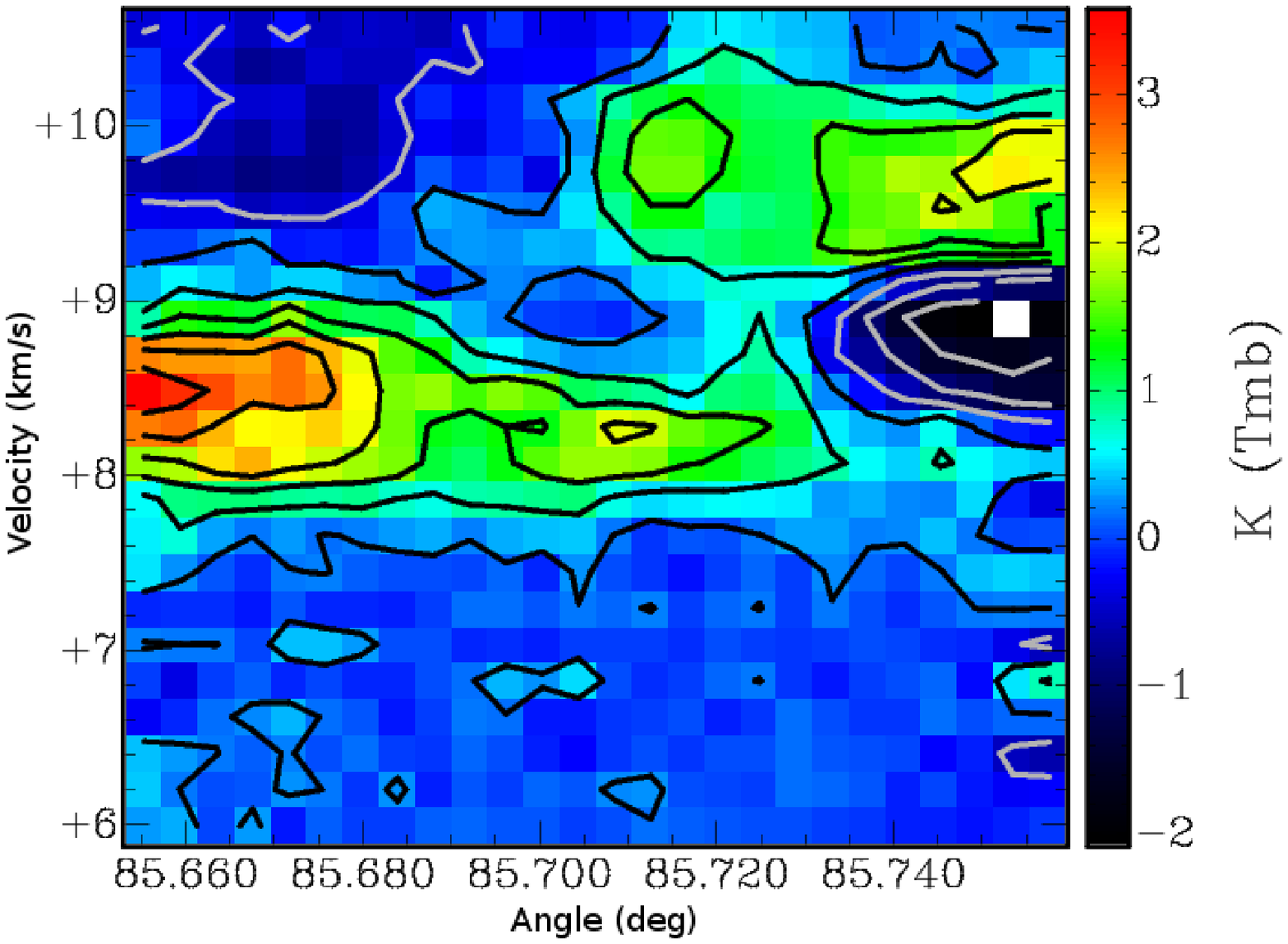}}
\caption{Position-velocity (PV) diagram of the $^{13}$CO$(2-1)$ emission 
towards the environment of SMM 3. The PV slice runs from NW to SE, across 
SMM 3, as shown in Fig.~\ref{figure:moments}. The black contours go from 10 to 
90\% of the maximum value of 3.6 K, in steps of 10\%. The grey contours 
represent negative values.}
\label{figure:pv}
\end{figure}

Figure~\ref{figure:pv} shows a position-velocity 
(PV) diagram for $^{13}$CO$(2-1)$ emission with the direction of the cut shown 
in Fig.~\ref{figure:moments}. The slice is taken through the entire mapped 
area, extending from the northwest corner to southeast corner, and its position 
angle, measured east of north, is P.A.$=135\degr$. As shown in the top middle 
panel of Fig.~\ref{figure:moments}, the PV slice goes across the border of the 
velocity gradient. The presence of two velocity components at 
$\sim8.5$ and $\sim9.5$ km~s$^{-1}$ are clearly visible in the PV plot. 
These velocities correspond to the NW and SE parts of the mapped area, 
respectively [the PV diagram of C$^{18}$O emission (not shown) is essentially 
similar].

\subsection{Spectra, line parameters, and column densities at the 
selected positions}

\begin{figure}[!h]
\centering
\resizebox{0.75\hsize}{!}{\includegraphics{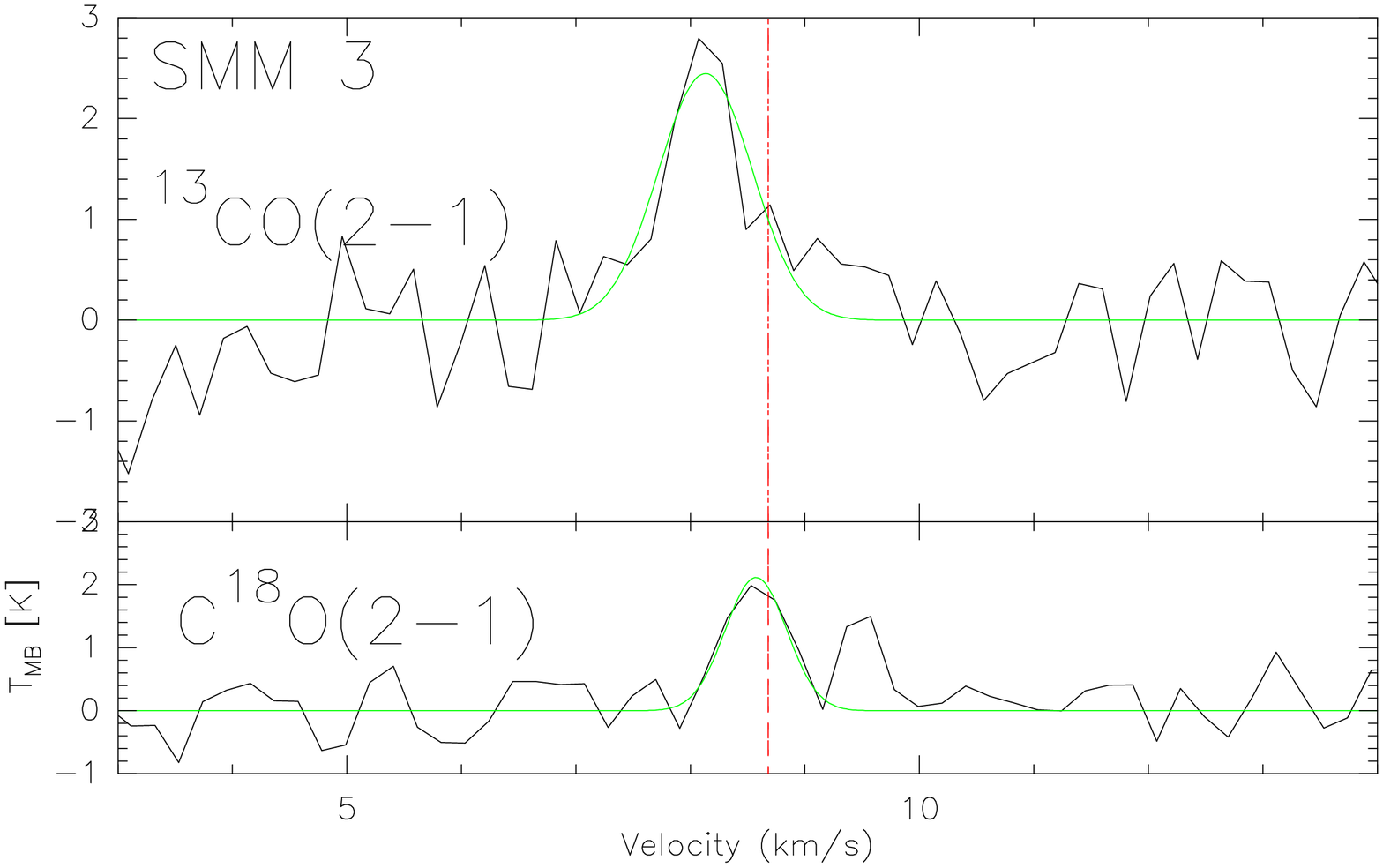}}
\resizebox{0.75\hsize}{!}{\includegraphics{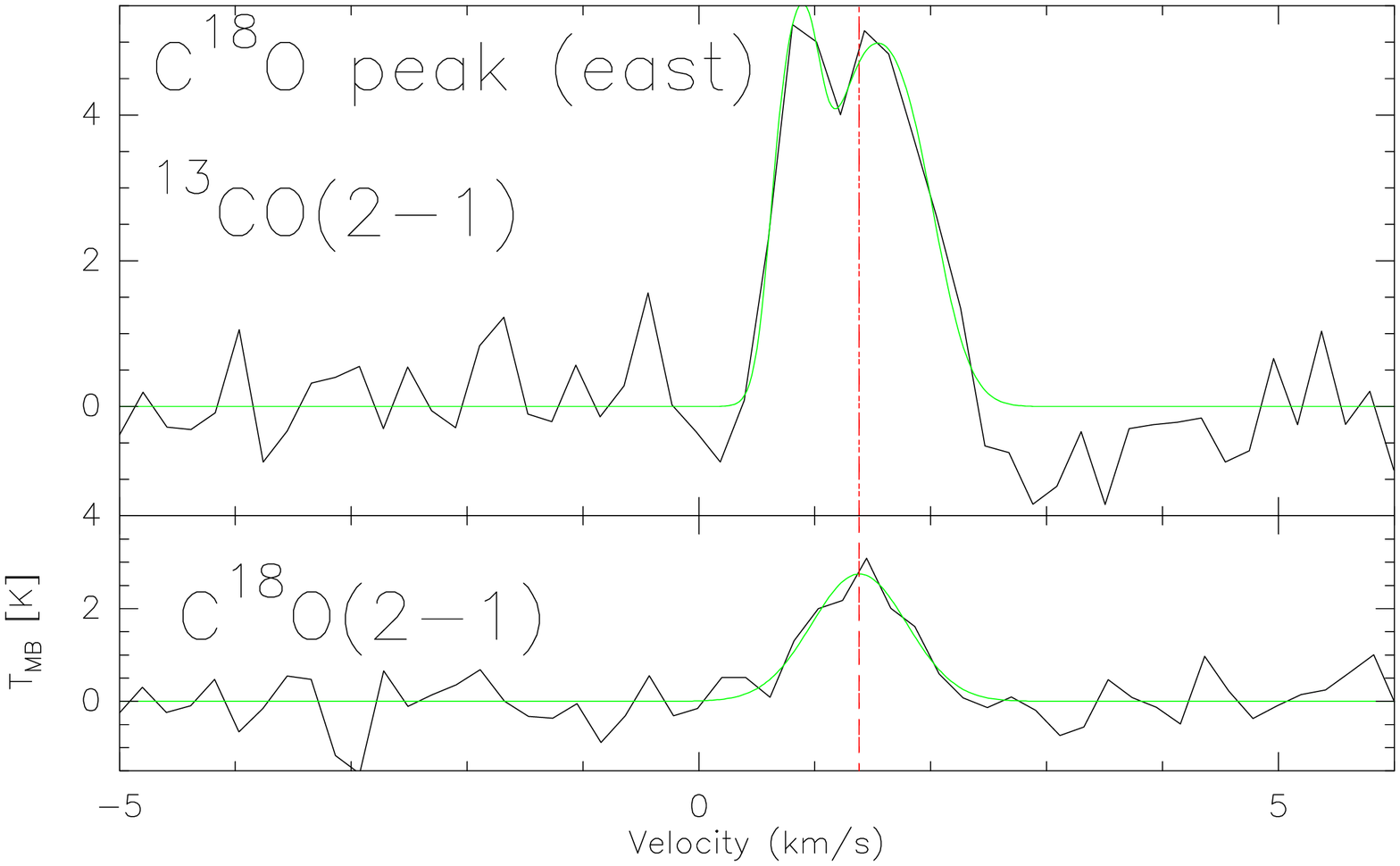}}
\caption{The $^{13}$CO$(2-1)$ and C$^{18}$O$(2-1)$ spectra extracted towards 
SMM 3 (\textit{upper panel}) and the position indicated in 
Fig.~\ref{figure:moments_second} (\textit{lower panel}). Hyperfine-structure 
fits to the $^{13}$CO lines, and single Gaussian fits to the C$^{18}$O lines 
are overlaid in green. The red dashed line overlaid on the spectra towards 
SMM 3 indicates the systemic velocity of SMM 3 as derived from 
C$^{17}$O$(2-1)$ in Paper III. In the lower panel, the red dashed 
line shows the radial velocity of the C$^{18}$O$(2-1)$ line.}
\label{figure:spectra}
\end{figure}

We extracted individual $^{13}$CO and C$^{18}$O spectra from the map 
centre, i.e., near the LABOCA 870-$\mu$m peak of SMM 3. Moreover, to 
investigate the properties of the lower-velocity component, we extracted the 
$^{13}$CO and C$^{18}$O spectra from a peak position of C$^{18}$O integrated 
intensity lying east of SMM 3 in projection (see the left panels of 
Fig.~\ref{figure:moments_second}). The extracted spectra are shown 
in Fig.~\ref{figure:spectra}. The C$^{18}$O line towards SMM 3 shows two 
nearby velocity components. There is a hint of that also in the $^{13}$CO 
spectrum. Moreover, the $^{13}$CO line appears to be blueshifted with respect 
to the C$^{18}$O line, suggesting that the lines originate in two
different gas layers along the line of sight. In the case of the lower 
velocity-component spectra (lower panel of Fig.~\ref{figure:spectra}), the 
$^{13}$CO line shows two velocity components around the systemic velocity 
of 1.38 km~s$^{-1}$ derived from C$^{18}$O$(2-1)$.

The line parameters of the extracted spectra are listed in 
Table~\ref{table:lineparameters}. In this table, we give the offset of the 
target position from the map centre (in arcsec), radial LSR velocity 
(${\rm v}_{\rm LSR}$), FWHM linewidth ($\Delta {\rm v}$), peak intensity 
($T_{\rm MB}$), integrated intensity ($\int T_{\rm MB} {\rm dv}$), peak line 
optical-thickness ($\tau_0$), and excitation temperature ($T_{\rm ex}$). 
The values of ${\rm v}_{\rm LSR}$ and $\Delta {\rm v}$ of the $^{13}$CO$(2-1)$ 
lines were derived through fitting the hyperfine structure of the line (see 
\cite{cazzoli2004}), while those of C$^{18}$O$(2-1)$ were obtained by fitting 
the lines with a single Gaussian. The two nearby velocity components 
were fitted simultaneously but for SMM 3 we only give the parameters 
of the components overlaid with green lines in 
Fig.~\ref{figure:spectra}. In contrast, in the case of the C$^{18}$O peak of 
the lower velocity-component, we list the $^{13}$CO parameters of the 
component slightly redshifted from the C$^{18}$O line velocity. 
The integrated line intensities listed in Col.~(7) of 
Table~\ref{table:lineparameters} were computed over the velocity range given 
in square brackets in the corresponding column. This way, we were 
able to take the non-Gaussian features of some of the lines into account. 
The quoted uncertainties in ${\rm v}_{\rm LSR}$ and $\Delta {\rm v}$ are formal 
$1\sigma$ fitting errors, while those in $T_{\rm MB}$ and 
$\int T_{\rm MB} {\rm dv}$ were estimated by summing in quadrature the fitting 
error and the 10\% calibration uncertainty. 

The values of $\tau_0$ and $T_{\rm ex}$ for the lines towards the 
C$^{18}$O peak of the lower velocity-component were derived as follows. 
By making the assumption that the $^{13}$CO and C$^{18}$O emission arise from 
the same gas\footnote{The LSR velocities of the two transitions are 
slightly different from each other [see Col.~(4) of 
Table~\ref{table:lineparameters}], so they may not be tracing exactly the same 
gas layers.} and the two transitions have the same beam filling factor 
and excitation temperature\footnote{We note that the observed $^{13}$CO and 
C$^{18}$O transitions have similar frequencies. Therefore, the 
frequency-dependent main-beam efficiency ($\eta_{\rm MB}$), and the telescope 
beam HPBW are also almost identical for the two transitions.}, we can 
numerically estimate the line optical thicknesses from the ratio of the line 
peak intensities as (e.g., \cite{myers1983}; \cite{ladd1998})

\begin{equation}
\frac{T_{\rm MB}({\rm ^{13}CO})}{T_{\rm MB}({\rm C^{18}O})}\approx \frac{1-e^{-\tau_{13}}}{1-e^{-\tau_{18}}}\,.
\end{equation}
The peak intensities are used here rather than the integrated 
intensities because multiple velocity components could ``contaminate'' the 
latter values. To calculate the transition optical thicknesses, we used the 
CO isotopologue abundance ratio of 

\begin{equation}
\frac{[{\rm ^{13}CO}]}{[{\rm C^{18}O}]}=\frac{[{\rm ^{13}C}]}{[{\rm ^{12}C}]}\frac{[{\rm ^{16}O}]}{[{\rm ^{18}O}]}=\frac{1}{60}\times500\simeq 8.3 \,.
\end{equation}
The adopted carbon- and oxygen-isotopic ratios are the same as those used in 
Paper III for a proper comparison to that work (see references 
therein)\footnote{We note that larger ratios of 
$[^{12}{\rm C}]/[^{13}{\rm C}]=77$ and $[^{16}{\rm O}]/[^{18}{\rm O}]=560$ 
(\cite{wilson1994}) are often used in a similar analysis (e.g., 
\cite{teyssier2002}). These values lead to the ratio $[^{13}{\rm CO}]/[{\rm C^{18}O}]\simeq7.3$.}. Consequently, the optical thickness ratio 
$\tau_{13}/\tau_{18}$ was assumed to be equal to 8.3. 

Once the optical thickness is determined, $T_{\rm ex}$ can be calculated using 
the radiative transfer equation [see, e.g., Eq.~(1) in Paper I]. In this 
calculation, we assumed a beam filling factor of unity, and that the 
background temperature is equal to the cosmic background radiation temperature 
of 2.725 K. 



The derived values of $\tau_{13}$, $\tau_{18}$, and $T_{\rm ex}$ are listed 
on Cols.~(8) and (9) of Table~\ref{table:lineparameters}. 
In this table, we use the symbol $\tau_0$ for the optical thickness to denote 
its peak value. The quoted uncertainties were derived from the uncertainties 
of the peak intensities. The $^{13}$CO line appears to be optically 
thick, while the C$^{18}$O line shows a moderate optical thickness of 
$\tau_{18}\simeq 0.8_{-0.2}^{+0.3}$. 

The total beam-averaged $^{13}$CO and C$^{18}$O column densities were computed 
following Eq.~(4) of Paper I. The spectroscopic parameters needed in 
the analysis, such as the electric dipole moments and rotational constants, 
were adopted from the JPL database. The derived column densities are listed in 
the last column of Table~\ref{table:lineparameters}. The quoted uncertainties 
were propagated from those associated with $T_{\rm ex}$, $\Delta {\rm v}$, 
and $\tau_0$ (the average value of the $^{+}_{-}-$errors of $T_{\rm ex}$ and 
$\tau_0$ were used).

\begin{table*}
\caption{Parameters of the spectra extracted from the selected positions.}
{\scriptsize
\begin{minipage}{2\columnwidth}
\centering
\renewcommand{\footnoterule}{}
\label{table:lineparameters}
\begin{tabular}{c c c c c c c c c c}
\hline\hline 
Position & Offset\tablefootmark{a} & Transition & ${\rm v}_{\rm LSR}$ & $\Delta {\rm v}$ & $T_{\rm MB}$ & $\int T_{\rm MB} {\rm dv}$\tablefootmark{b} & $\tau_0$ & $T_{\rm ex}$ & $N$\\
   & $(\arcsec,\, \arcsec)$& & [km~s$^{-1}$] & [km~s$^{-1}$] & [K] & [K~km~s$^{-1}$] & & [K] & [$10^{15}$ cm$^{-2}$]\\
\hline
{\bf Main velocity component}\\
SMM 3 & $(0\,, 0)$ & $^{13}$CO$(2-1)$ & $8.11\pm0.09$ & $0.95\pm0.36$ & $2.73\pm0.45$ & $2.61\pm0.67$ [7.17, 9.21] & - & - & -\\
      & & C$^{18}$O$(2-1)$ & $8.57\pm0.05$ & $0.63\pm0.11$ & $2.12\pm0.25$ & $1.39\pm0.25$ [8.00, 9.14] & $0.2\pm0.1$\tablefootmark{c} & $11.0\pm1.0$\tablefootmark{c} & $0.85\pm 0.2$\tablefootmark{d}\\
{\bf Lower-velocity component}\\
C$^{18}$O peak & $(+85.5\,, +14.0)$ & $^{13}$CO$(2-1)$ & $1.52\pm0.06$ & $0.81\pm0.12$ & $5.17\pm0.61$ & $7.05\pm0.86$ [0.21, 2.38] & $6.3^{+2.4}_{-1.9}$\tablefootmark{e} & - & $18.2\pm6.8$\\
     & & C$^{18}$O$(2-1)$ & $1.38\pm0.06$ & $0.97\pm0.15$ & $2.75\pm0.61$ & $2.89\pm0.47$ [0.07, 2.52] & $0.8^{+0.3}_{-0.2}$\tablefootmark{e} & $9.5\pm1.8$\tablefootmark{f} & $2.8\pm1.0$\\
\hline 
\end{tabular} 
\tablefoot{
\tablefoottext{a}{Offset from the map centre in 
arcsec.}\tablefoottext{b}{Integrated intensity is derived by integrating over 
the velocity range indicated in square brackets.}\tablefoottext{c}{This optical thickness is estimated from that derived for C$^{17}$O$(2-1)$ in Paper III, and the reported $T_{\rm ex}$ value is that of C$^{17}$O$(2-1)$.}\tablefoottext{d}{Derived from the integrated intensity under the assumption of optically thin emission.}\tablefoottext{e}{Derived from the line intensity ratio, as described in Sect.~3.4.}\tablefoottext{f}{Derived using the $T_{\rm MB}$ and $\tau_0$ values of the C$^{18}$O line.}}
\end{minipage} }
\end{table*}

\subsection{CO depletion in SMM 3}

In Paper III, we derived a CO depletion factor of $f_{\rm D}=10.8\pm2.2$ 
towards SMM 3 through C$^{17}$O$(2-1)$ observations ($27\farcs8$ 
re\-solution). Another estimate of $f_{\rm D}$ in SMM 3 can be 
obtained from the current C$^{18}$O data. We prefer to use C$^{18}$O 
for this analysis rather than $^{13}$CO, because C$^{18}$O emission is more 
optically thin than the $^{13}$CO emission. Therefore, C$^{18}$O is expected 
to trace gas deeper into the core's envelope, and being less affected by 
foreground emission.

The values of $\tau_0$ and $T_{\rm ex}$ for the C$^{17}$O$(2-1)$ line 
were derived to be $0.05\pm0.03$ and $11.0\pm1.0$ K, respectively 
(Paper III). As the oxygen-isotopic ratio $[^{18}{\rm O}]/[^{17}{\rm O}]$ is 
only 3.52 (\cite{frerking1982}), it can be assumed that the C$^{18}$O$(2-1)$ 
line is also optically thin. Under the assumption of optically thin emission 
($\tau \ll 1$), and adopting $T_{\rm ex}=11\pm1$ K, the C$^{18}$O column 
density computed from the integrated intensity is 
$N({\rm C^{18}O})=8.5\pm 1.7 \times 10^{14}$ cm$^{-2}$.

By smoothing the LABOCA map to match the resolution of the C$^{18}$O map 
($30\arcsec$), and regridding it onto the same pixel scale, the 870-$\mu$m 
peak flux density towards the $(0\arcsec,\, 0\arcsec)$ position of the 
C$^{18}$O map is determined to be 0.85 Jy~beam$^{-1}$. Using the assumption 
that $T_{\rm dust}=T_{\rm kin}$, and that the dust opacity per unit dust mass
at 870 $\mu$m is $\kappa_{870}=1.7$ cm$^2$~g$^{-1}$ (\cite{ossenkopf1994}; 
Paper I), we estimate that the corresponding H$_2$ column density is 
$N({\rm H_2})=2.5\pm0.3\times10^{22}$ cm$^{-2}$ (see Papers I and III for 
further details). Therefore, the fractional abundance of C$^{18}$O towards the 
$(0\arcsec,\, 0\arcsec)$ position is estimated to be 
$x({\rm C^{18}O})=N({\rm C^{18}O})/N({\rm H_2})=3.4\pm0.8\times10^{-8}$.

To compute $f_{\rm D}$ from C$^{18}$O data, we need an estimate of the 
``canonical'', or undepleted, abundance of C$^{18}$O. Using the standard value 
$9.5\times10^{-5}$ for the abundance of the main CO isotopologue in the solar 
neighbourhood (\cite{frerking1982}), we can write 

\begin{equation}
x({\rm C^{18}O})_{\rm can}=x({\rm CO})_{\rm can}\times \frac{[{\rm ^{18}O}]}{[{\rm ^{16}O}]}=9.5\times10^{-5}\times \frac{1}{500}=1.9\times10^{-7}\,.
\end{equation}
The value of $f_{\rm D}$ is then determined to be 
$f_{\rm D}=x({\rm C^{18}O})_{\rm can}/x({\rm C^{18}O})_{\rm obs}=5.6\pm1.3$. 
This is comparable within a factor of two to the value derived 
from C$^{17}$O$(2-1)$ data. The agreement is quite good given all 
the assumptions used in the analysis. We note that the 
$(0\arcsec,\, 0\arcsec)$ position of the C$^{18}$O map is not exactly 
coincident with our previous C$^{17}$O observation target position 
but the two are within the beam size of both the observations.



Following the analysis presented in Miettinen (2012; Sect.~5.5 
therein), the CO depletion timescale in SMM 3 is estimated to be only 
$\tau_{\rm dep}\sim1-3.7\times10^4$ yr [using the values $T_{\rm kin}=11.3$ K 
and $n({\rm H_2})=1.1-4.0\times10^5$ cm$^{-3}$; see Table~\ref{table:SMM3}]. 
This provides a lower limit to the age of the core.

\section{Discussion}

\subsection{On the non-detection of molecular outflows}

Because SMM 3 is an early Class 0 source, it is expected to drive a 
bipolar molecular outflow. The outflows can ma\-nifest themselves in broad 
non-Gaussian spectral-line wings. One of our original attempts of the present 
study was to search for outflows driven by SMM 3 through $^{13}$CO 
observations. However, in our data, there is no evidence for a large-scale 
$^{13}$CO outflow. In the \textit{Spitzer}/IRAC 4.5-$\mu$m image of 
SMM 3 (Fig.~\ref{figure:SMM3}; bottom panel), there are some 4.5-$\mu$m 
emission features that could be signatures of shock-excited material around 
SMM 3 (e.g., \cite{smith2005}; \cite{debuizer2010}). For comparison, the 
Class 0/I protostar IRAS 05399-0121 in Orion B9, which drives the 
HH 92 jet (\cite{bally2002}), exhibits spectacular IRAC 4.5-$\mu$m features 
along its parsec-scale bipolar jet. Clearly, higher resolution observations, 
and better outflow tracers, such as $^{12}$CO and SiO, would be needed to 
clarify the outflow activity of SMM 3.


\subsection{Low-velocity gas emission}

As was discussed in Paper II (Sect.~5.7 therein), the lower-velocity 
line emission seen towards Orion B9 is likely to come from the 
``low-velocity part'' of Orion B, which probably ori\-ginates from the 
feedback from the massive stars of the nearby Ori OB 1b association. 
This fraction of the gas is likely to be located a few tens of 
parsecs closer to the Sun than the ``re\-gular'' 9-km~s$^{-1}$ gas component 
(\cite{wilson2005}).

The C$^{18}$O column density towards the selected C$^{18}$O peak 
is $2.8\pm1.0\times10^{15}$ cm$^{-2}$. We can convert this to an estimate of 
the H$_2$ column density as 

\begin{equation}
N({\rm H_2})=\frac{[{\rm H_2}]}{[{\rm ^{12}CO}]}\times \frac{[{\rm ^{12}CO}]}{[{\rm C^{18}O}]}\times N({\rm C^{18}O}) \,.
\end{equation}
Using again the values 
$[{\rm ^{12}CO}]/[{\rm C^{18}O}]=[{\rm ^{16}O}]/[{\rm ^{18}O}]=500$ and 
$[{\rm ^{12}CO}]/[{\rm H_2}]=9.5\times10^{-5}$, we obtain 
$N({\rm H_2})=1.5\pm0.5\times10^{22}$ cm$^{-2}$. This shows that the 
``low-velocity part'' of Orion B also consists of dense gas, which conforms to 
our previous detection of, e.g., deuterated molecular species at comparable 
LSR velocities. We also note that the prestellar core SMM 7 and Class 
0 protostar IRAS 05413-0104 seen towards Orion B9 have such low systemic 
velocities ($\sim3.7$ and $\sim1.5$ km~s$^{-1}$, respectively) that they are 
likely to be members of the low-velocity Orion B. Despite the estimated high 
column density of the clump-like $^{13}$CO/C$^{18}$O feature seen in the left 
panels of Fig.~\ref{figure:moments_second}, it was not seen in 
LABOCA 870-$\mu$m emission at the $\sim3\sigma$ level (reflecting the 
strong CO depletion in the main velocity component).

\subsection{On the origin of the velocity gradient and implications 
for the core/star formation in Orion B9}

In Paper II, we speculated that the Orion B9 region has probably been 
influenced by the feedback from the nearby Orion OB association, or more 
precisely, from the Ori OB 1b subgroup. As mentioned above, this 
feedback is believed to be responsible for the ``low-velocity part'' 
of Orion B (\cite{wilson2005}). We believe that the 
discovery of a velocity gradient in the present study supports the possibility 
that Orion B9 region is affected by stellar feedback. 

\begin{figure*}
\begin{center}
\includegraphics[scale=0.7]{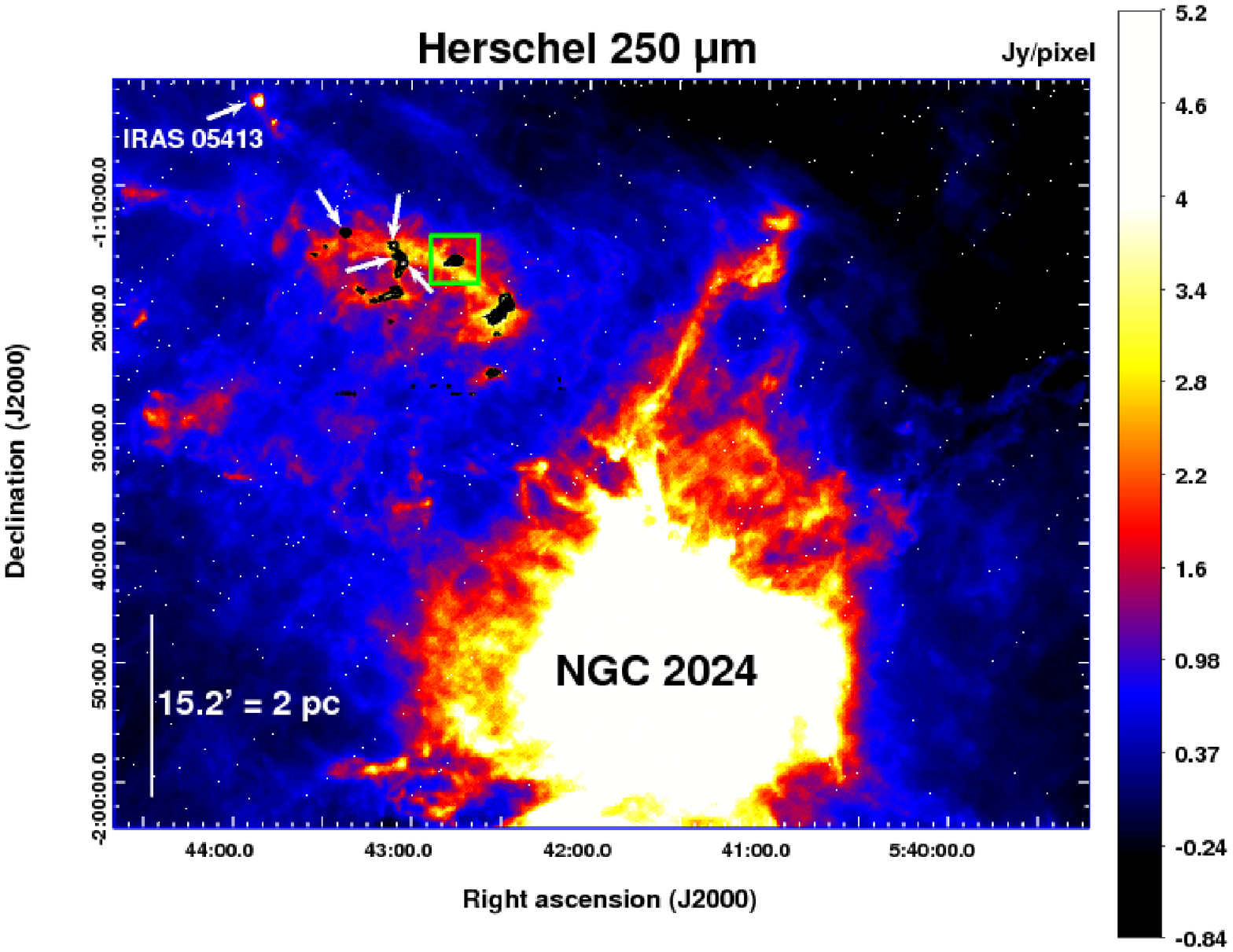}
\includegraphics[scale=0.6]{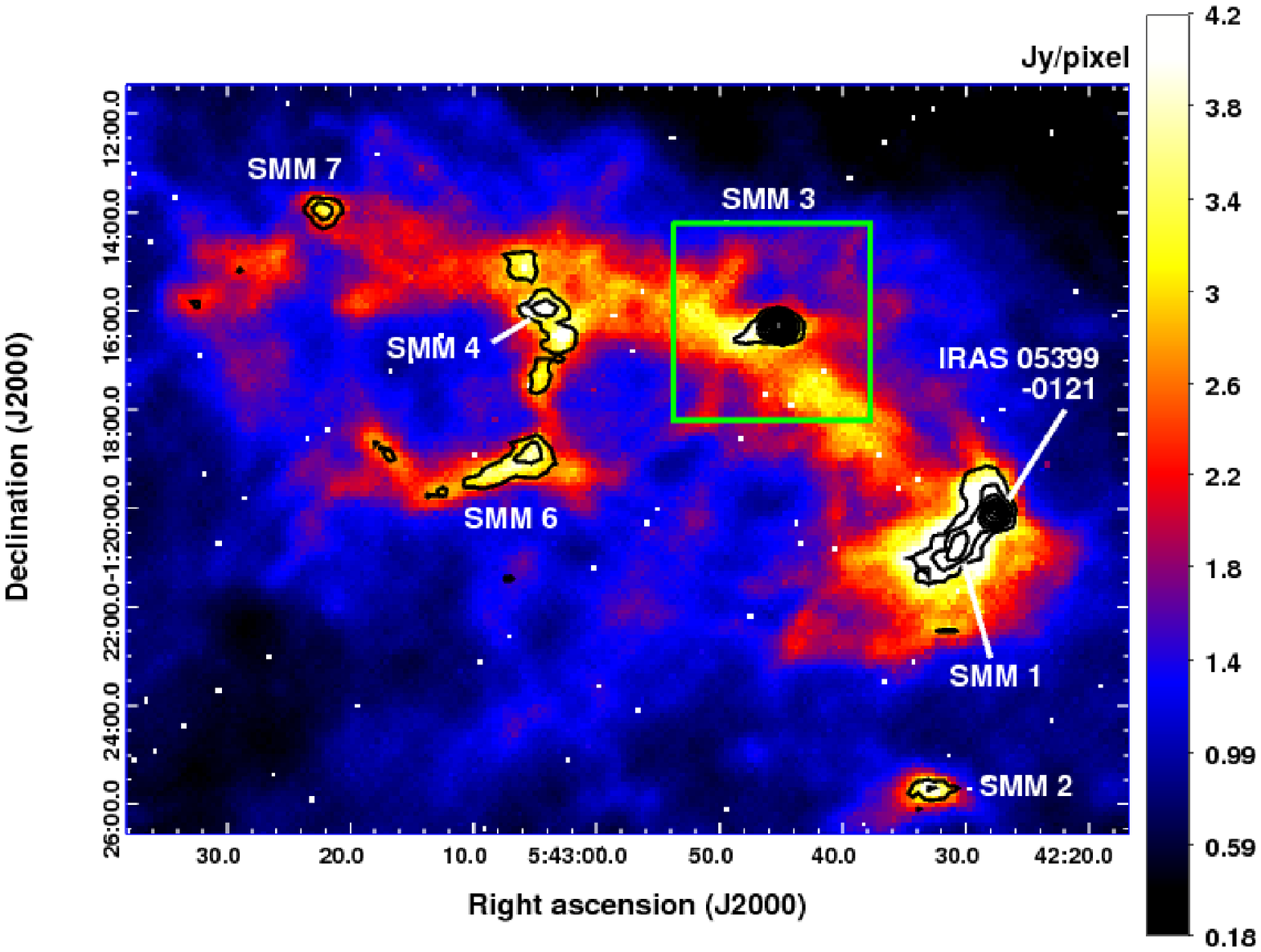}
\caption{\textbf{Top:} A wide-field \textit{Herschel}/SPIRE 250-$\mu$m 
far-infrared image towards the Orion B9 and NGC 2024 (Flame Nebula) 
star-forming regions. The image is overlaid with black contours of 
LABOCA 870-$\mu$m dust continuum emission. 
The contours go from 0.1 ($\sim3.3\sigma$) to 1.0 Jy~beam$^{-1}$ in 
steps of 0.1 Jy~beam$^{-1}$. The green rectangle outlines the 
$4\arcmin \times 4\arcmin$ area mapped in $^{13}$CO$(2-1)$ and 
C$^{18}$O$(2-1)$ in the present study. The white arrows indicate the cores 
which have a lower LSR velocity than the ``main 9-km~s$^{-1}$ part'' of 
Orion B9 or show multiple velocity components. The Class 0 protostar 
IRAS 05413-0104 in the northeast corner was at the border of our LABOCA map 
and only weakly detected (see Fig.~1 in Paper I). Note how the Orion B9 cores 
are associated with a NE-SW oriented filamentary structure, and the 
cores with lower radial velocity (or with multiple velocity components) lie at 
the NE part of the structure. A scale bar indicating the 2 pc projected length 
is shown in the bottom left, with the assumption of a 450 pc line-of-sight 
distance. \textbf{Bottom:} A zoomed-in view of the upper image towards Orion 
B9. Selected cores are labelled. The LABOCA contours are as in the 
upper panel.}
\label{figure:spire}
\end{center}
\end{figure*}

To better illustrate the larger-scale view of the surroundings of Orion B9, in 
Fig.~\ref{figure:spire} we show a wide-field \textit{Herschel}/SPIRE 
250-$\mu$m image\footnote{\textit{Herschel} is an ESA space observatory with 
science instruments provided by European-led Principal Investigator consortia 
and with important participation from NASA (\cite{pilbratt2010}). Orion B was 
observed as part of the ``\textit{Herschel} Gould Belt Survey (GBS)'' 
(\cite{andre2010}), using the PACS (\cite{poglitsch2010}) and the SPIRE 
(\cite{griffin2010}) instruments. For more details, see 
{\tt http://gouldbelt-herschel.cea.fr}. The data are available from the 
Herschel Science Archive (HSA) at 
{\tt http://herschel.esac.esa.int/Science$_{-}$Archive.shtml}}. 
A visual inspection of the image suggests that the Orion B9 region might 
be situated in quite a dynamic environment. Towards the south, 
there is the active massive star-forming region NGC 2024 some $40\arcmin$ 
($\sim5.2$ pc at 450 pc) from Orion B9. The northwest-southeast 
oriented 250-$\mu$m filament emanating from NGC 2024 is likely related to the 
dense molecular ridge of NGC 2024 running in the same direction 
(e.g., \cite{thronson1984}; \cite{visser1998}; \cite{watanabe2008}). 
The expanding \ion{H}{ii} region of NGC 2024 appears to 
be interacting with the molecular ridge (\cite{gaume1992}). The majority of 
the members of Ori OB 1b association lie towards the west/northwest from Orion 
B9 (see, e.g., Fig.~1 in \cite{wilson2005}). 

It is apparent from the bottom panel of Fig.~\ref{figure:spire} that 
the Orion B9 cores, with the possible exception of SMM 7, belong to 
a common northeast-southwest oriented filamentary structure. As 
indicated in the fi\-gure, the cores at the northeastern part of the region 
have a lower LSR velocity or show multiple velocity components as found in our 
previous papers. As discussed earlier, a gradient of increasing radial 
velocity from NW to SE is seen across the mapped area, and SMM 3 appears to 
lie on the border of the velocity jump [cf.~the case of the Class 0 protostar 
HH211 in Perseus/IC348 (\cite{tobin2011})]. Interestingly, this velocity 
gradient appears to be orthogonal to the direction of the 250-$\mu$m filament 
oriented NE-SW (SMM 3 also lies on the border of the 
dust filament). This raises the question whether this border could be tracing 
a shock layer of the interacting/colliding flows, within which there 
is a jump in the velocity of the gas. Such interaction might have 
been responsible for triggering the formation of SMM 3, and (some) of the 
other cores in the region. It seems more likely that, instead of the influence 
of Ori OB 1b, the feedback from the nearby expanding NGC 2024 \ion{H}{ii} 
region could have compressed the inital cloud region (\cite{fukuda2000}). 
Later, the cloud under a high pressure gradient may have fragmented into dense 
cores, out of which some, such as SMM 3, were collapsed into protostars 
sequentially. 

SMM 3 showed the highest level of CO depletion among the cores studied in 
Paper III. This conforms to the fact that it also appears to be the densest 
core in Orion B9. 
For the core collapse induced by compression, the simulations by Hennebelle et 
al. (2003) suggest that the combined duration of the prestellar$+$Class 0 
phase is $\sim3.2\times10^5-1.3\times10^6$ yr (depending on the rate of 
compression). As mentioned earlier, the estimated CO depletion timescale in 
SMM 3 is $\sim1-3.7\times10^4$ yr, which is shorter than the core evolution 
timescales quoted above. On the other hand, the fragmentation timescale of the 
core is expected to be comparable to the signal crossing time, which for SMM 3 
is estimated to be $\tau_{\rm cross}=D/\sigma_{\rm 3D}\sim6.6\times10^5$ yr (the 
projected core diameter across the $3.3\sigma$ contour is $86\arcsec$ or 
$D=0.19$ pc, and $\sigma_{\rm 3D}=\sqrt{3}\sigma_{\rm 1D}$ is the 
three-dimensional velocity dispersion). This agrees well with the above
theore\-tical core lifetimes. From these considerations, we suggest 
that the formation of SMM 3, and of some other cores in Orion B9, was 
triggered by feedback from NGC 2024 (dynamical compression) some several 
times $\sim10^5$ yr ago (cf.~\cite{fukuda2000}).

To better understand the velocity structure of the region on larger scales, 
larger maps of molecular-line emission would be needed. We note that the star 
formation in Orion B9, if triggered by stellar feedback, might resemble the 
situations in the $\rho$ Ophiuchus (e.g., \cite{nutter2006}) and the B59/Pipe 
Nebula (\cite{peretto2012}), where the star formation appears to be induced by 
the feedback from the Scorpius OB association.

\section{Summary and conclusions}

A $4\arcmin \times 4\arcmin$ region around the Class 0 protostar SMM 3 in 
Orion B9 was mapped in $^{13}$CO and C$^{18}$O $J=2-1$ lines with the APEX 12-m 
telescope. Our main results and conclusions can be summarised as follows:

\begin{enumerate}
\item Both lines exhibit two well separated velocity components: one at 
$\sim1.3$ km~s$^{-1}$ and the other at $\sim8.7$ km~s$^{-1}$. The latter is 
near the systemic velocity of SMM 3. The low-velocity component was already 
recognised in our previous studies, and it is believed to be related to the 
low-velocity part of Orion B.
\item The $^{13}$CO and C$^{18}$O emission are relatively widely distributed 
compared to the dust continuum emission traced by LABOCA. The LABOCA 
870-$\mu$m peak position of SMM 3 is not coincident with any strong $^{13}$CO 
or C$^{18}$O emission, which is in accordance with the high CO depletion 
factor derived earlier by us from C$^{17}$O$(2-1)$ ($f_{\rm D}\simeq10.8$). 
The CO depletion factor derived from C$^{18}$O data is within a factor of two 
from the previous estimate, i.e., $f_{\rm D}\simeq5.6$. No evi\-dence 
for a large-scale outflow activity, i.e., high velocity line wings, was found 
towards SMM 3.
\item The lower-velocity ($\sim1.3$ km~s$^{-1}$) $^{13}$CO and C$^{18}$O 
emission are concentrated into a clump-like feature at the eastern part of the 
map. We estimate that the H$_2$ column density towards its C$^{18}$O maximum 
is $\sim10^{22}$ cm$^{-2}$. Therefore, the lower-velocity gas seen along the 
line of sight is of high density, which is consistent with our earlier 
detection of, e.g., deute\-rated molecular species (DCO$^+$, N$_2$D$^+$).
\item We observe a velocity gradient across the $^{13}$CO and C$^{18}$O maps 
along the NW-SE direction (some hint of that is also visible in the 
lower-velocity line maps). Interestingly, SMM 3 is projected almost exactly on 
the border of the velo\-city jump. The sharp velocity-gradient border 
provides a strong indication that it represents an interaction zone of flow 
motions. 
\item We suggest a possible scenario in which the formation of SMM 3, and 
likely some of the other dense cores in Orion B9, was triggered by 
an expanding \ion{H}{ii} region of NGC 2024. This collect-and-collapse -type 
process might have been taken place some several times $10^5$ yr ago. 
The NGC 2024 region is known to be a potential site of induced, 
sequential star formation (e.g., \cite{fukuda2000}, and references therein). 
The case of Orion B9 suggests that we may be witnessing the most recent event 
of self-propagating star formation around NGC 2024. Larger-scale 
molecular-line maps would be needed for a better understanding of the 
larger-scale velocity structure of the region.
\end{enumerate}


\begin{acknowledgements}

I thank the anonymous referee very much for his careful reading and 
constructive comments and suggestions which helped to improve this paper 
considerably. I am grateful to the staff at the APEX telescope for performing 
the service-mode observations presented in this paper. The Academy of Finland 
is acknowledged for the financial support through grant 132291. 
SPIRE has been developed by a consortium of institutes led by Cardiff Univ. 
(UK) and including: Univ. Lethbridge (Canada); NAOC (China); CEA, LAM 
(France); IFSI, Univ. Padua (Italy); IAC (Spain); Stockholm Observatory 
(Sweden); Imperial College London, RAL, UCLMSSL, UKATC, Univ. Sussex (UK); 
and Caltech, JPL, NHSC, Univ. Colorado (USA). This development has been 
supported by national funding agencies: CSA (Canada); NAOC (China); CEA, 
CNES, CNRS (France); ASI (Italy); MCINN (Spain); SNSB (Sweden); STFC, UKSA 
(UK); and NASA (USA).
This research has made use of NASA's Astrophysics Data System and the 
NASA/IPAC Infrared Science Archive, which is operated by the JPL, California 
Institute of Technology, under contract with the NASA. This research has also 
made use of the SIMBAD database, operated at CDS, Strasbourg, France.

\end{acknowledgements}

\end{document}